\documentclass[twocolumn]{aastex7}

\usepackage[dvipsnames]{xcolor}
\usepackage{acronym}
\usepackage{todonotes}
\usepackage{rotating}
\usepackage{enumitem}
\usepackage{amsmath}
\usepackage{booktabs}

\definecolor{Mulberry}{HTML}{A93C93}
\definecolor{JungleGreen}{HTML}{00A99A}
\definecolor{Amethyst}{HTML}{9966CC}
\definecolor{Aquamarine}{HTML}{7FFFD0}
\definecolor{Capri}{HTML}{00BFFF}

\newcommand{\bkj}[1]{\textcolor{magenta}{{\bf #1}}}

\submitjournal{ApJ}

\graphicspath{{./}{figures/}}

\begin{document}




\title{Historical Reconstruction of Solar Surface Magnetism from Cycle~1\,--\,24 Using the Synthetic Active Region Generator (SARG) and the Advective Flux Transport (AFT) Model}

\author[0000-0003-3191-4625]{Bibhuti Kumar Jha}
\affiliation{Southwest Research Institute, Boulder, CO 80302, USA}
\email{bkjha.sun@gmail.com}

\author[0000-0003-0621-4803]{Lisa A. Upton}
\affiliation{Southwest Research Institute, Boulder, CO 80302, USA}
\email{lisa.upton@swri.org}

\author{Greg  Kopp}
\affiliation{University of Colorado, Laboratory for Atmospheric and Space Physics, 3665 Discovery Drive, Boulder, CO 80303, USA}
\email{greg.kopp@lasp.colorado.edu}

\author[0000-0002-4338-7028]{Odele Coddington}
\affiliation{University of Colorado, Laboratory for Atmospheric and Space Physics, 3665 Discovery Drive, Boulder, CO 80303, USA}
\email{Odele.Coddington@lasp.colorado.edu}

\shorttitle{AFT 2.0: Historical Reconstruction}
\shortauthors{B. K. Jha et al.}
\correspondingauthor{Bibhuti Kumar Jha}
\email{bkjha.sun@gmail.com}



\begin{abstract}
The historical reconstruction of the Sun’s surface magnetic field remains a persistent challenge, limiting our ability to investigate the long-term global properties of the Sun, including the evolution of the large-scale magnetic field, solar cycle prediction, reconstruction of total solar irradiance (TSI), and secular solar variability. In this study, we employ the Advective Flux Transport (AFT) model in conjunction with our newly developed Synthetic Active Region Generator (SARG) to construct a catalog of synthetic active regions (ARs) spanning Solar Cycles~1\,--\,24 (1755\,--\,2020). We use the SIDC/SILSO sunspot number version~2.0 as the sole input governing the properties of the synthetic ARs in this catalog. This SARG catalog is then incorporated into the AFT model, which simulates the emergence of new ARs on the Sun, which are then transported under the influence of surface flows to produce maps of the full-Sun radial photospheric magnetic field over the entire 265-year period. We modulate the active region tilt for each cycle in order to ensure that the polar fields are consistent with the solar cycle amplitudes. We find that the polar fields derived from these simulations exhibit excellent correlation (\(r > 0.8\)) with observational proxies, including polar faculae counts and Ca\,\textsc{ii}~K polar network indices. Daily synchronic maps from these simulations for the entire 265-year period are made publicly available to support a wide range of applications beyond those presented in this work.
   
\end{abstract}

\keywords{Solar active regions(1974) --- Sunspots(1653) --- Solar cycle(1487) --- Bipolar sunspot groups(156)}


\acrodef{sftm}[SFTM]{Surface Flux Transport Model}
\acrodef{sft}[SFT]{Surface Flux Transport}
\acrodef{hfam}[FAM]{Far-side Acoustic Map}
\acrodef{pinn}[PINN]{Physics-Informed Neural Network}
\acrodef{ml}[ML]{Machine Learning}
\acrodef{cme}[CME]{coronal mass ejection}
\acrodef{ar}[AR]{active region}
\acrodef{bmr}[BMR]{bipolar magnetic region}
\acrodef{nrt}[NRT]{near-real-time}
\acrodef{sdo}[SDO]{Solar Dynamics Observatory}
\acrodef{soho}[SOHO]{Solar and Heliospheric Observatory}
\acrodef{hmi}[HMI]{Helioseismic and Magnetic Imager}
\acrodef{mdi}[MDI]{Michelson Doppler Imager}
\acrodef{solis}[SOLIS]{Synoptic Optical Long-term Investigations of the Sun}
\acrodef{sidc}[SIDC]{Solar Influences Data Analysis Center }
\acrodef{silso}[SILSO]{Sunspot Index and Long-term Solar Observations}
\acrodef{phi}[PHI]{Polarimetric and Helioseismic Imager}
\acrodef{so}[SO]{Solar Orbiter}
\acrodef{stereo}[STEREO]{Solar Terrestrial Relations Observatory}
\acrodef{euvi}[SECCHI-EUVI]{Sun Earth Connection Coronal \& Heliospheric Investigation-Extreme Ultraviolet Imager}
\acrodef{euv}[EUV]{extreme ultraviolet }
\acrodef{gong}[GONG]{Global Oscillation Network Group}
\acrodef{los}[LOS]{line-of-sight}
\acrodef{jsoc}[JSOC]{Joint Science Operations Center}
\acrodef{cr}[CR]{Carrington Rotation}
\acrodef{urm}[URM]{under-represented minorities}
\acrodef{aftm}[AFT]{Advective Flux Transport}
\acrodef{adapt}[ADAPT]{Air Force Data Assimilative Photospheric Flux Transport}
\acrodef{oft}[OFT]{Open Flux Transport}
\acrodef{hipft}[HipFT]{High-Performance Flux Transport}
\acrodef{reu}[REU]{Research Experience for Undergraduate students}
\acrodef{pmi}[PMI]{Photospheric Magnetic field Imager}
\acrodef{pde}[PDE]{Partial Differential Equation}
\acrodef{fno}[FNO]{Fourier Neural Operator}
\acrodef{dnn}[DNN]{Deep Neural Network}
\acrodef{cnn}[CNN]{Convolutional Neural Networks}
\acrodef{aft}[AFT]{Advective Flux Transport}
\acrodef{aft2}[AFT\,2.0]{Advective Flux Transport Version 2.0}
\acrodef{dft}[DFT]{discrete Fourier transform}
\acrodef{ace}[ACE]{Advanced Composition Explorer}
\acrodef{dscovr}[DSCOVR]{Deep Space Climate Observatory}
\acrodef{tsi}[TSI]{Total Solar Irradiance}

\section{Introduction} \label{sec:intro}

The discovery of the Sun’s magnetic field in the early 20th century - through Zeeman splitting of spectral lines - was a landmark in solar physics, revealing magnetism as a fundamental driver of solar dynamics \citep{Hale1908, Hale1919}. While the first full-disk magnetograms were obtained in the 1950s \citep{Babcock1953}, systematic synoptic observations of the photospheric magnetic field did not begin until the 1970s \citep{Howard1974, Livingston1976}. The absence of full-disk magnetic measurements prior to this era presents a critical challenge for understanding long-term solar variability \citep{Karak2023}, predicting solar cycles, modeling heliospheric evolution, and reconstructing total solar irradiance (TSI). To estimate surface magnetism before direct observations were available, significant efforts have focused on reconstructing the Sun’s historical magnetic field using empirical models and indirect proxies \citep{Jiang2011,Pevtsov2019,Virtanen2019a,Virtanen2022}. The \ac{sft} model \citep{Sheeley1992, Jiang2014a, Yeates2023} has become an essential tool for such reconstruction efforts.

The \ac{sft} model is a widely used tool for simulating the evolution of the Sun's photospheric magnetic field, which is driven by large-scale plasma flows in the Sun's convective region. \ac{sft} models enable the construction of full-Sun synchronic magnetic maps guided by observations indicative of magnetic activity. By modeling the entire solar surface, \ac{sft} models provide an avenue for filling in the observational limitations caused by only viewing one hemisphere of the Sun from Earth at a time and by limb regions suffering from line-of-sight effects \citep{hickmann2015,Upton2014, Hathaway2016, Caplan2025}. Such full-surface knowledge has been critical for space weather forecasting and coronal modeling.

To extend photospheric magnetic field simulations prior to regular full-disk magnetic-field measurements,  \ac{sft} models have used limited sunspot and plage (Ca II K) observations \citep{Jiang2011, Virtanen2019a, Virtanen2022, Yeates2025} to incorporate magnetic sources. These simulations rely on assumptions derived from the known statistical properties of solar \acp{ar}, such as Hale`s polarity law, Joy's law \citep{Hale1908, Hale1919}, to fill in the gaps. While sunspot and plage observations are more limited in temporal coverage, sunspot counts spanning Solar Cycles 1–24 (1755\,--\,2020;) provide over two centuries of information about AR emergence, a data source to construct the historical magnetic filed going back to early 17th century.

A successful reconstruction, over this time period offers a powerful framework to investigate multi-century solar cycle variability enabling new insights into the Sun’s role in long-term space weather and Earth-climate variability. Additionally, this is directly and indirectly affected by solar irradiance changes\citep{Greg2016}, which, in turn, are highly correlated to solar magnetic activity \citep[see review by][]{Gray2010}. This raises a fundamental question: \textit{Can we create a continuous reconstruction of the solar photospheric magnetic field over a period as long as 24 solar cycles (1755–2020) using only the sunspot number and statistical properties of \ac{ar}s?}

A few previous efforts, such as \citet{Schrijver2002, Wang2005, Jiang2011, Wang2021}, have demonstrated the possibility of using sunspot number as the only input into an SFT model to reconstruct the historical magnetic field of the Sun. Although these prior works laid the foundation of such efforts, they were limited by the available observations. In particular, the data was insufficient to provide statistically quantified characteristics of solar ARs, forcing them to make assumptions. However, \citet{Jiang2011a} was able to create an observationally well-informed synthetic AR catalogs to reconstruct the historical magnetic field. Unfortunately, these catalogs still lack some fundamental observational properties, e.g., ARs emergence rate, stochasticity in the nature of flux emergence etc. We have developed a code, the Synthetic Active Region Generator \citep[SARG:][]{Jha2024a}, for creating more observationally constrained catalogs with multiple realizations to accurately capture the stochastic nature of flux emergence. Using these AR catalogs as input with one of the most realistic SFT models (AFT), to then produced high resolution magnetograms ($\approx0.35^\circ$/pixel, with 512$\times$1024 pixels, in latitude and longitude) going back to 1755.


In this work, we reconstruct the solar photospheric magnetic field over the past 265 years (Cycles 1–24) using the sunspot number (SSN v2.0) as the sole observational input by combining the novel SARG catalog with the \acl{aft} \citep[AFT:][]{Upton2014, Hathaway2016} model. SARG uses known statistical properties of \acp{ar} to create a synthetic \ac{ar} catalog. The  \acp{ar} in this catalog are then used as input for the \ac{aft}, which simulates the emergence of \acp{ar} on the solar surface and their subsequent evolution. By combining these tools, we generate full-Sun radial magnetic field maps that are consistent with known active region emergence patterns and large-scale transport processes. This synthetic reconstruction not only reproduces key solar cycle features, such as the reversal and buildup of polar fields, but also enables quantitative comparisons with historical proxies of polar magnetism. The resulting set of magnetic maps provides a physically consistent and observationally constrained view of the Sun’s surface magnetic evolution over more than two centuries.

In the following sections, first we briefly describe the \ac{aft} model, as well as recent improvements that have been included. We refer to this improved version as the  \ac{aft2} model. Secondly, we discuss the underlying algorithm of SARG, followed by the methodology that we use to incorporate the \acp{ar}  into the model. Finally, we discuss our resulting historical reconstruction.

\section{Advective Flux Transport Model}\label{sec:AFT2}

\ac{sft} models solve the magnetohydrodynamic (MHD) transport equation to simulate the evolution of the Sun’s photospheric radial magnetic field, creating maps of the radial magnetic field \(B_r\) over the entire surface of the Sun. The fundamental equation governing the evolution of \(B_r\) under the influence of advection and diffusion, often called advection-diffusion equation, is given by
\begin{equation}
    \frac{\partial B_r}{\partial t} + \vec{\nabla} \cdot (\vec{u}B_r) - \eta \nabla^2 B_r = S(\theta, \phi, t).
\label{eq1}
\end{equation}

Here, $\vec{u}$ represents the horizontal velocity field, which includes axisymmetric components such as differential rotation and meridional flow, as well as convective flows \citep{Hathaway2011, RightmireUpton2012, Upton2014, Upton2014a}, while $\eta$ and $S$ denote the diffusivity and the source term, respectively, the latter representing the emergence of new magnetic flux at the solar surface.

Unlike traditional SFT models, the \ac{aft} model and the more recent OFT model \citep{Caplan2025} incorporate simulated convective flows (supergranule scales and larger) to more realistically represent the plasma flows observed on the Sun. This explicit convective approach distinguishes these models from many earlier SFT implementations that relied on a constant turbulent diffusivity to approximate the effects of convection \citep{Hathaway2011, RightmireUpton2012}. In AFT, the convective flows are simulated using time-evolving vector spherical harmonics, capturing the observed spectral characteristics and lifetimes of large-scale convective cells. These convective cells are then advected by the measured differential rotation and meridional flow. AFT also includes a field-dependent flow attenuation for plasma velocities in magnetically active regions. This weakening of the flows is representative of the change in plasma $\beta$ as the field strength increases - an effect that is more difficult to capture with a constant diffusivity and that is often overlooked in SFT models. The diffusion term \(\eta \nabla^2 B_r\) in Equation~\ref{eq1} is included solely for numerical stability and does not significantly influence the transport of magnetic flux \citep[see][]{Upton2014, Upton2014a}.

\subsection{AFT\,2.0}

\ac{aft2} is an improved version of \ac{aft} that leverages advances in computational power and modern multiprocessing capabilities, which significantly reduces the computational time needed to complete a simulation. This is critical for conducting \ac{sft} simulations that span many solar cycles (in our case: 24 cycles spanning more than 250 years). This improvement in computational time also enables the ability to quickly run dozens of cases (ensembles) for better uncertainty quantification \cite{Jha2024a}. Redesigned as a flexible framework, \ac{aft2} has the ability to incorporate the source term, $S(\theta, \phi, t)$ (\autoref{eq1}) in three different modes of operation:

\begin{itemize}[left=0.01pt]
    \item AFT-DA: In this mode (\textit{a.k.a.}, the AFT Baseline Mode), AFT performs data assimilation of magnetograms to construct a full-synchronic magnetic map of the Sun \citep{Upton2014, Upton2014a}. The model solves the induction equation over the entire surface, but then uses weights to combine the simulation with the observations. This allows the model to find a solution where data is not available while correcting for any deviations in locations where observations are available. In this mode, the magnetograms serve as the source term for the model. Currently, magnetograms from \ac{soho}/\ac{mdi} (AFT-DA.MDI) and \ac{sdo}/\ac{hmi} (AFT-DA.HMI) are used. 

    \item AFT-FS: In this mode, AFT incorporates far-side \acp{ar} from observational proxies in addition to the standard near-side observations, enabling smoother updates to the source term by accounting for  \ac{ar}  emergence on the far side of the Sun. This allows the model to capture a more complete and robust representation of the Sun’s surface magnetic field. Specifically, \ac{aft2} utilizes far-side proxies of \ac{ar} from helioseismic and EUV sources. STEREO 304\,\AA\ \citep[AFT-FS.304][]{Upton2024A}, \ac{hmi} far-side helioseismology (AFT-FS.HMI), and GONG far-side helioseismology (AFT-FS.GONG) have currently been adapted for use in \ac{aft2}.

    \item AFT-AR: In this mode, AFT incorporates \acp{ar} in the form of bipolar Gaussian spots based on observed properties (latitude, longitude, area, tilt, and magnetic flux) of the \acp{ar}. This mode can be used with the \acp{ar} observed in magnetograms (AFT-AR.OBS), but more often the \acp{ar} are prescribed in a real or synthetic \ac{ar} catalog (AFT-AR.SYNTH). This mode is particularly effective for simulating the evolution of large-scale magnetic field to make solar cycle prediction \citep{Jha2024a}.   

\end{itemize}



In this paper, we focus on the version of AFT that uses SARG (AFT-AR.SYNTH) to construct a time series of sunspot emergence based on the \ac{sidc}/\ac{silso} monthly smoothed sunspot number for the last 24 solar cycles, i.e., the period of 1755\,--\,2020.

\section{SARG}\label{sec:sarg}
\begin{figure}[htbp!]
    \centering
    \includegraphics[width=0.9\columnwidth]{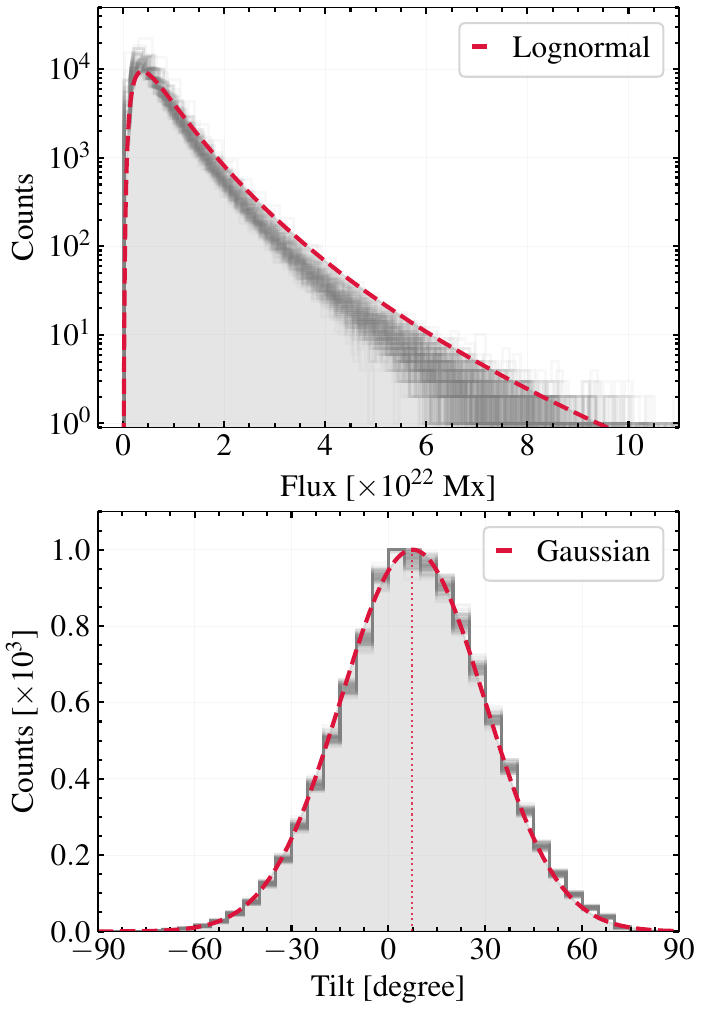}
    \caption{Distribution of magnetic flux (top) and tilt (bottom) for each synthetic AR across 30 SARG realizations. The red dashed curve represent the input empirical distribution of the magnetic flux and tilt of of \acp{ar}.}
    \label{fig:flux_distl}
\end{figure}

SARG creates a catalog (or catalogs) comprised of a list of \acp{ar}, each containing the time of emergence ($t_{\rm AR}$), latitudes and longitudes ($\theta,\phi$) of both the leading and following polarities, and the total unsigned flux ($\Phi_{\rm M}$) in the AR. The only input that SARG needs is a time series representing the monthly  sunspot number. In this study, we rely on the 13-month smoothed sunspot number series provided by \ac{sidc}/\ac{silso} (SSN~V2.0\footnote{WDC-SILSO, Royal Observatory of Belgium, Brussels; \citet{ssn}.}). This time series contains only numbers of sunspots as a function of time; it has no positional or size information. Via the SARG, the series is used as the basis for estimating the properties of \acp{ar} (e.g., emergence rate, latitude of the emergence, etc.). A similar approach has been used in \citet{Schrijver2001,Schrijver2002, Wang2005, Jiang2011}. The synthetic properties are derived by randomly sampling from the observed statistical properties of \acp{ar}. This allows countless realizations of \ac{ar} catalogs to be created from a single time series. The detailed steps SARG uses for generating synthetic \acp{ar} catalogs is outlined as follows:

\begin{enumerate}[left=0.1pt]
    \item At any given time ($t_{\rm AR}$), we start by randomly selecting the active region's magnetic flux ($\Phi_{\rm M}$) drawn from a log-normal flux distribution (see \autoref{fig:flux_distl}) derived in \citet{Munoj2015} from observations using the \ac{solis} instrument at the Kitt Peak Vacuum Telescope. This distribution is given by  
    $$\Phi_{\rm M} = \exp(\mu + \sigma \, X),$$  with $\mu=50.05$ and $\sigma=0.75$. Here, $X$ is a standard normal deviate $\mathcal{N}(0, 1)$ representing the observed skewness toward large-flux ARs.  See the entry for ``\textit{BMR Flux KPVT/SOLIS}" in Table 1 of \citet{Munoj2015}.  

    \item The migration of the mean latitude of emergence for \acp{ar} as a function of the solar-cycle phase is known as Sp\"{o}rer’s Law. In SARG, $\lambda (t)$ at $t_{\rm AR}$ is derived from two equations: one describing the centroid of the active region zone of emergence ($\mu_\lambda$) and one describing the width of the zone of \ac{ar} emergence ($\sigma_\lambda$). The central latitude is a function from the time of the start of the cycle \citep{Hathaway2011a} and is given by 
    \begin{align*}
    \mu_\lambda(t) &= \lambda_0 \exp\left(-\frac{t-t_0}{\tau}\right), ~\lambda_0=28^\circ~\&~\tau=90;\\
    \end{align*}
    where $t-t_0$ is the time in months from the start of the cycle. 
    The width of the zone of \ac{ar} emergence is a function of the total sunspot area in $\mu$Hem. For SIDC/SILSO (SSN~V1.0 \footnote{WDC-SILSO, Royal Observatory of Belgium, Brussels; \citet{ssn}.}), the total sunspot area in $\mu$Hem is given as $A=16.7\times$SSN~V1.0 \citep{Hathaway2015}. The primary conversion from SSN~V1.0 to SSN~V2.0 is a multiplicative factor (known as the Z{\"u}rich scale factor) of 0.6, such that $SSN~V1.0 \approx 0.6*SSN~V2.0$ \citep{Clette2016}. Folding this into the area relationship, we have   $A=10.0\times$SSN~V2.0.
    \citet[see Equation~9]{Hathaway2015} showed that the width of the zone of \ac{ar} emergence is given by 
    \begin{align*}
    \sigma_\lambda(A) = 1.5^\circ + 3.8^\circ \left[ 1 - \exp\left( \frac{-A_{\rm Total}}{400} \right) \right]
    \end{align*}
    where $A_{\rm Total}$ is the total sunspot area in $\mu$Hem (calculated from the sunspot number at $t_{\rm AR}$). For each \ac{ar}, SARG chooses the latitude of emergence from a random distribution $\mathcal{N}(\mu_\lambda, \sigma_\lambda)$. 
    This distribution, as a function of time and cycle amplitude, mimics the latitudinal migration of the active region belts, forming the characteristic ``butterfly wings'' that solar cycle progression resembles. In \autoref{app:butterfly}, \autoref{fig:sarg_butter} we show the time-latitude butterfly diagram of four SARG realizations.  
    
    
    \item  SARG assumes that \acp{ar} emerge randomly across all longitudes. 
    Thus, SARG samples the longitude ($\phi$) of each \ac{ar} from a random uniform distribution across $\phi \in [0^\circ, 360^\circ)$. No attempt is made to simulate possible active longitudes and their durations.

    \item Once $\lambda$, $\phi$ and $\Phi_{\rm M}$ are selected for the \ac{ar}, SARG determines the longitudinal separation based on the size of the \ac{ar}. First the \ac{ar} flux is used to estimate the sunspot area (A) in $\mu$Hem \citep{Upton2014, Upton2024A} using the relation $\Phi_{\rm M}=7\times10^{19}A$ \citep{Sheeley1966,  Mosher1977}. The longitudinal separation is then given by \cite{Upton2024A} $$\Delta \phi = 3.0^\circ + 8.0^\circ \tanh\left[\frac{A}{500}\right].$$

    \item The mean tilt ($\gamma$) of an \ac{ar} is assigned using the standard Joy's law equation \citep{Hale1919}, $$\gamma= \mathcal{N}(\mu_\lambda,\sigma=22.31),$$ where $\mu_\lambda=32.0^\circ \sin{\lambda}$ \citep{Stenflo2012a}. SARG adds a stochastic randomness on top of this standard Joy's law tilt to match with the observed tilt distribution \citep[see \autoref{fig:flux_distl} and, ][]{MunozJaramillo2021}. 
    The latitudinal separation is calculated from the tilt and the $\Delta\phi$, $$\Delta\lambda=\Delta\phi\cos{\lambda}\tan{\gamma}.$$

    \item Finally, the rate of AR emergence is stochastically calculated from the monthly sunspot number \citep{Upton2025_are}, with the waiting time between successive emergence $\Delta t$ calculated as
    \begin{equation*}
        \Delta t = \frac{365.25/12}{a + b\left[{\rm ISSN }+\mathcal{N}(0, 5)\right]},
    \end{equation*}
    where $a=0.3$\,day$^{-1}$ and $b=0.2697$\,day$^{-1}$ are empirical parameters calibrated to reproduce observed AR emergence rates.

    We derive these best-fit parameters ($a$ and $b$) by empirically fitting the relation between the time lag ($\Delta t$) --- the time gap between the emergence of two consecutive AR groups observed on the visible hemisphere of the Sun --- and the monthly smoothed SSN v2.0, using least-square fit. To get $\Delta t$, we search in a sunspot group catalog that was created by calibrating and merging data from the Royal Observatory of Greenwich an the Solar Optical Observing Network/US National Oceanic and Atmospheric Administration \citep{hathaway_2025_17108109}. This sunspot catalog covers the period of 1874\,--\,2025. We start by identifying all the unique sunspot groups and then calculate the time lag between new consecutive AR groups as a function of cycle strength. Detailed analysis is presented in the upcoming article \citet{Upton2025_are}. These steps are repeated in steps of $t+\Delta t$ until the desired date and time is reached. 
\end{enumerate}

We generate a SARG AR catalog that includes key parameters for each \ac{ar} as described above. It is important to note that due to the stochastic nature of the SARG algorithm, no two realizations produce identical sets of active regions. However, all realizations are statistically consistent, sharing the same underlying distributions for key \ac{ar} properties. In \autoref{fig:flux_distl}, we show the distributions of absolute magnetic flux, \(\Phi_{\rm M}\), and tilt angle, \(\gamma\), for 100 different SARG realizations. The figure demonstrates that, while individual realizations vary, the overall statistical behavior remains consistent across the realizations.


\section{Methodology}

\begin{figure}[!htbp]
    \centering
    \includegraphics[width=0.9\columnwidth]{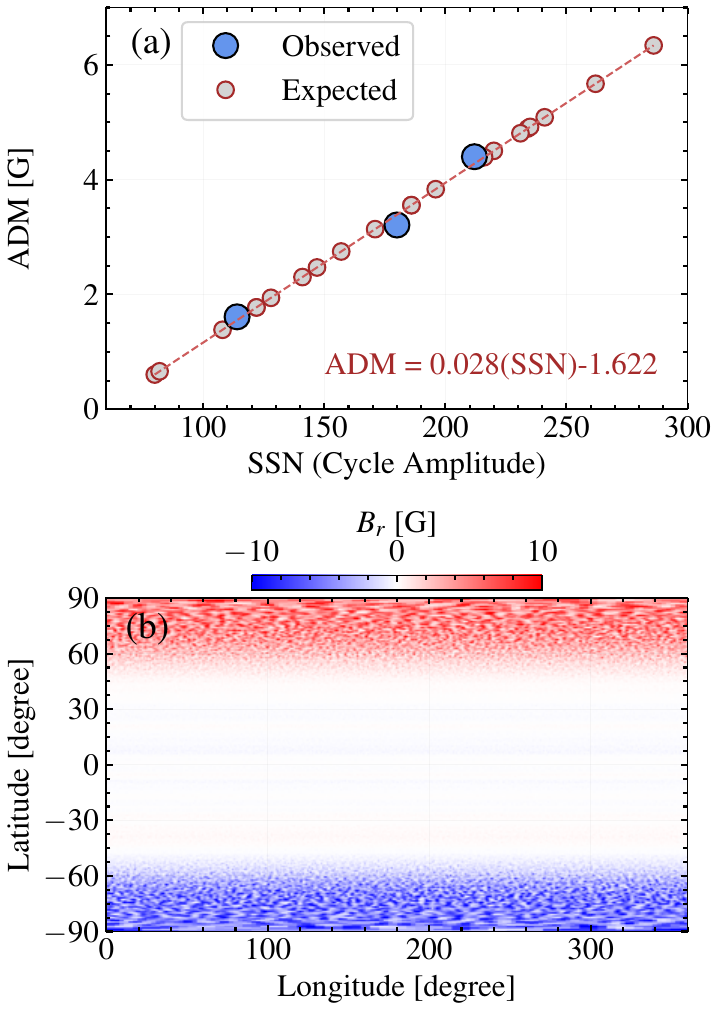}
    \caption{(a) The relationship between ADM at the solar cycle minimum and strength of the following solar cycle. Here the observed ADM is the dipole component calculated from the WSO observations (blue), whereas the estimated ADM values (red) are interpolated/extrapolated using the best fit straight line  (dashed) through observed data points. These expected values are the values we target by adjusting the tilt. (b) The initial dipole configuration used in the model at the beginning of Solar Cycle\,1.}
    \label{fig:admssnl}
\end{figure}

The Sun's axial dipole moment (ADM), a measure of the strength of the global magnetic field at the time of solar minimum, correlates with the amplitude of the upcoming solar cycle. With this expected correlation, we derive an empirical relationship between the ADM and the strength of the subsequent solar cycle based on WSO observations over three solar cycles \citep[represented by a filled circles in \autoref{fig:admssnl}a;][]{Upton2023}. A linear fit to these data provides an estimate of the strength of the ADM for each cycle minimum. The best fit line estimated based on the least square fitting is ${\rm ADM} = 0.028 {\rm (SSN)}-1.622$.

Our simulation begins in 1755, at the onset of Solar Cycle~1, therefore we use this linear fit to estimate the strength of the ADM at the time of the solar minimum immediately prior to Solar Cycle 1. This estimate is then used to create a map of the photospheric radial magnetic field to serve as a starting magnetic field configuration with a perfect dipolar ADM configuration to initialize the AFT historical reconstruction.  In \autoref{fig:admssnl}b, we show the initial \(B_r(T = 0)\) map used at the beginning of the simulation period. This represents the magnetic-field configuration 


We then run AFT using the SARG \ac{ar} catalog as input to AFT. We  begin by creating a map on a Carrington Heliographic grid ($\lambda-\phi$) with idealized bipolar regions, with each bipole having a two dimensional Gaussian shape, for each of the \acp{ar} scheduled to emerge on that day.

%


\subsection{Modification in SARG tilt distribution}

The strength of Sun's dipolar magnetic field at solar cycle minimum sets the amplitude of the following cycle and therefore serves as one of the most important constraints on solar cycle simulations. However, there are little to no direct observations of the Sun's magnetic field prior to the modern era. This limits our ability to validate the results of the model simulations. Therefore, we also use the linear fit described above to guide the historical reconstruction of the solar cycle evolution. This framework serves as a cross-check that allows us to ensure that polar magnetic fields created during each cycle are consistent with the observational constraints. 

To accomplish this, we systematically adjust the tilts for all of the \acp{ar} (starting from Cycle~1) in steps of $0.5^\circ$ for a range of modified tilt values $\Delta \gamma \in [-5^\circ,5^\circ]$ (effectively shifting the mean of the tilt distribution to the left or right, see \autoref{app:tiltmod}, \autoref{fig:sarg_modtilt}). This results in an ensemble of varying-tilt SARG-catalog ARs. For a given number of sunspots, ensembles with more inclined tilts are capable or causing faster cancellation of the ADM near solar maximum and subsequently a stronger ADM buildup for the following solar minimum. Selection of tilt adjustment (via different-tilt ensembles) allows matching of successive solar cycles via an iterative approach.

\begin{figure}[htbp!]
    \centering
    \includegraphics[width=\columnwidth]{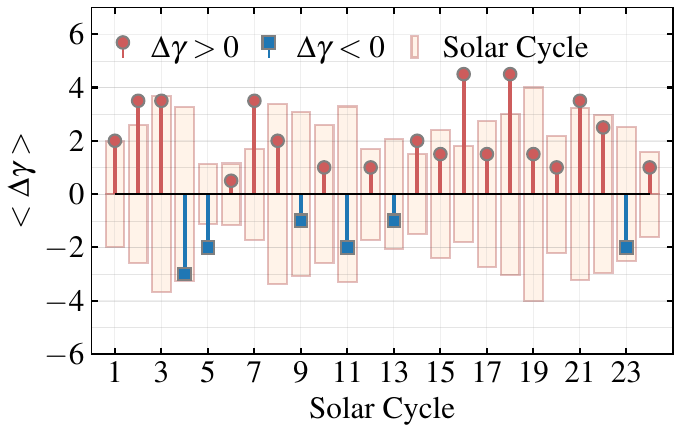}
    \caption{Distribution of $<\Delta\gamma>$, applied in each solar cycle to obtain the desired ADM at the end of the cycle. The shaded background colored bar show the relative strength of the solar cycle (mirrored on both side of x-axis for better comparison).}
    \label{fig:modtilt}
\end{figure}

We begin by running the AFT simulation using the default Joy's tilt prescribed in the SARG catalog. We then run the AFT simulation using tilt-modified SARG catalogs and select the simulation in which produces a simulated ADM at the end of each cycle that falls within a predefined tolerance ($\pm0.5$\,G). The resulting radial magnetic field map ($B_r$) at the end of each cycle is then used as the initial condition for the subsequent cycle, and the process is repeated. This iterative tuning process is successively applied to all solar cycles starting from Cycle~1 through then middle of Cycle~24 (the year 2020). A similar process was used by \citet{Wang1989, Wang1991a} by modifying the amplitude of the meridional circulation rather than modifying the tilt distribution. 
In \autoref{fig:modtilt} we show the modification in tilt required for each cycle to achieve the desired ADM at the end of the given solar cycle. An example of effect of tilt modification in the distribution of tilt is is also shown in \autoref{app:tiltmod} (\autoref{fig:sarg_modtilt}).

\section{Results}
\begin{figure}[!htbp]
    \centering
    \includegraphics[width=\columnwidth]{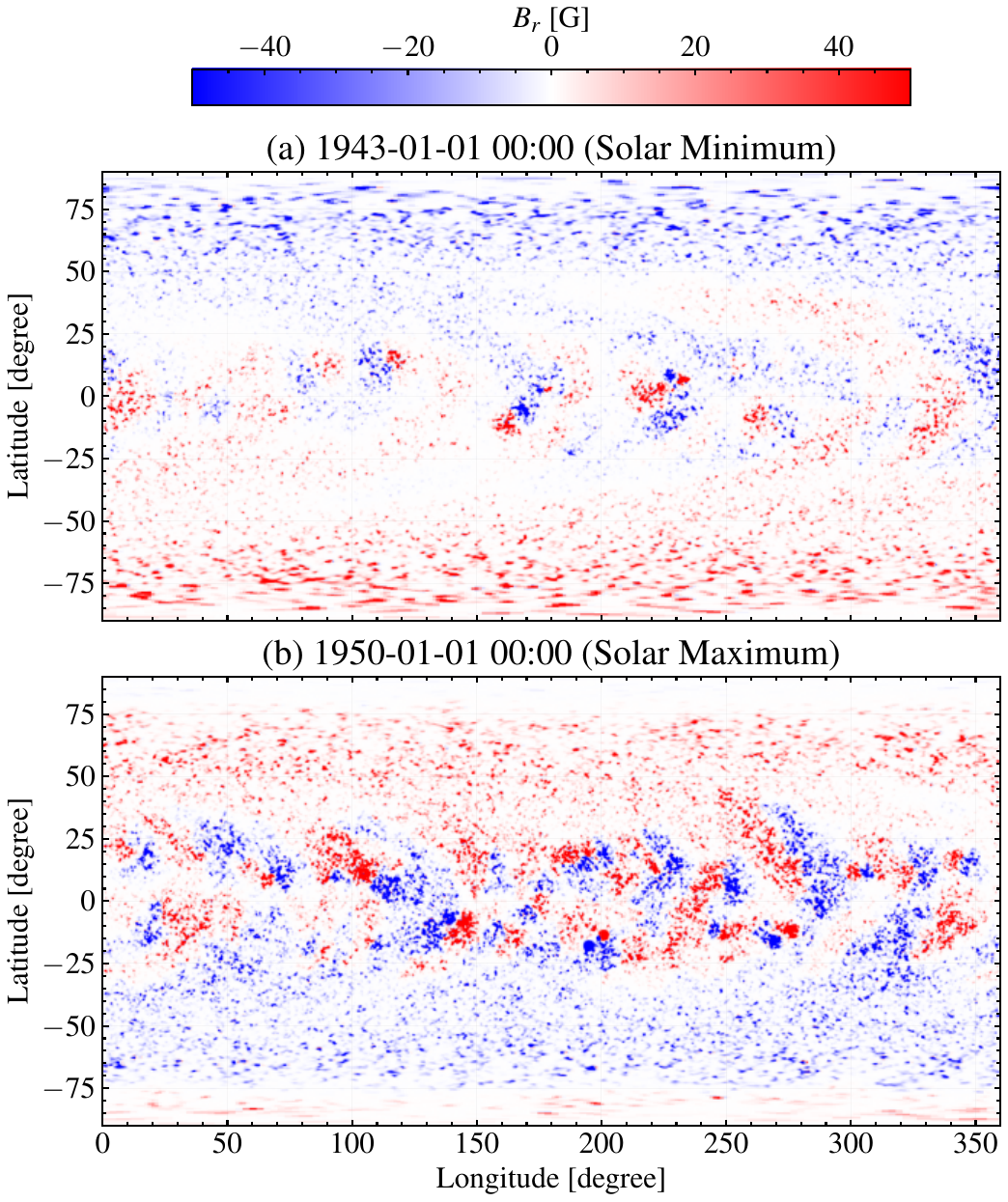}
    \caption{Two representative photospheric $B_r$ maps from Solar Cycle\,19 during a (a) low- and (b) high-activity period.}
    \label{fig:aftmap_ex}
\end{figure}

Adapting to the approach outlined above, we have simulated the photospheric evolution of the Sun's magnetic field for the period 1755\,--\,2020. In \autoref{fig:aftmap_ex}, we show example maps of the reconstructed photospheric radial magnetic field for two periods, a lower activity period and a higher activity period.. Such maps, created for the entire time range, enable historical reconstructions of the magnetic butterfly diagram, polar cap field strengths, and the ADM. 

\subsection{Global Magnetic Properties for last 24 solar cycles}

\begin{sidewaysfigure*}[htbp!]
    \includegraphics[width=0.96\textheight]{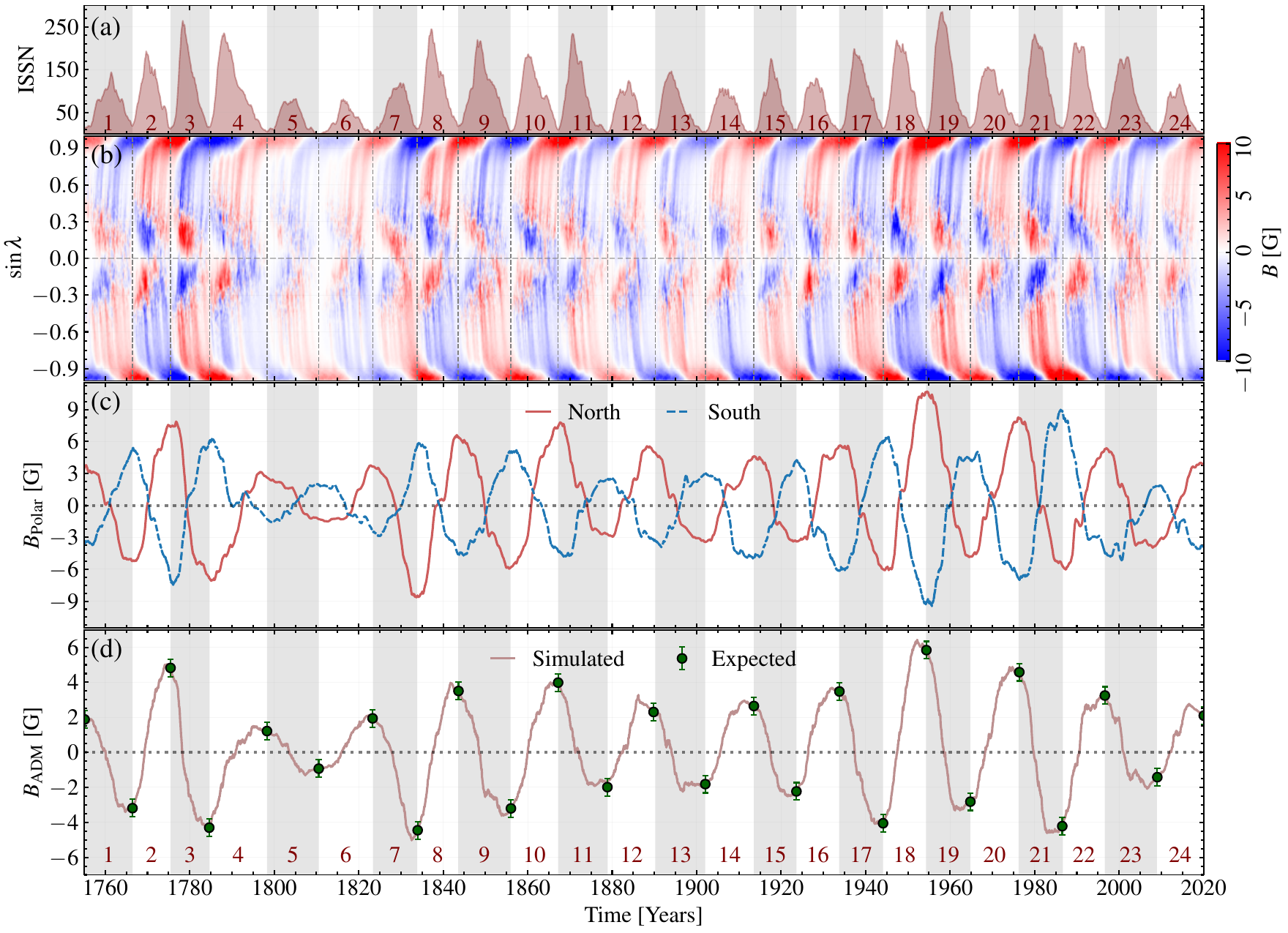}
    \caption{Reconstruction of the photospheric magnetic field over 24 solar cycles and the derived large-scale characteristics with tuned and no-zero $\Delta \gamma$ to match with estimated ADM at the end of the given solar cycle. (a) International 13-month smoothed sunspot number, (b) magnetic butterfly diagram, (c) The polar field calculated over latitude 60$^\circ$ and (d) the simulated ADM values along with that estimated from the empirical relation given in \autoref{fig:admssnl}.}
    \label{fig:tsi_all}
\end{sidewaysfigure*}

We constructed the magnetic butterfly diagram from the AFT-generated \(B_r\) maps using the conventional method of azimuthal averaging followed by averaging over each Carrington rotation (\(\approx 27.3\) days). This magnetic butterfly diagram, along with the monthly sunspot number variation (\autoref{fig:tsi_all}a), is shown in \autoref{fig:tsi_all}b. As evident in \autoref{fig:tsi_all}b, several key features of the observed solar cycle emerge. These features are characteristics that are hallmarks of our current understanding of solar cycle behavior \citep{Hathaway2015, Charbonneau2010} and solar dynamo theory \citep{Charbonneau2014}, thus reinforcing the robustness of our approach to reconstructing \(B_r\) over such an extended period.

The most apparent feature is the similarity in the shape and character of the butterfly wings. They possess the chevron-like shape that make up the wings. They have the expected Hale polarity in each wing and in each hemisphere. The shape and intensity of the wings change from one cycle to the next, with the stronger cycles extending to higher latitudes and having a distinctly stronger magnetic field strengths. 

Note particularly the alignment of the timing of the polar field reversal with the solar maximum, as seen in \autoref{fig:tsi_all}b and \ref{fig:tsi_all}c. This is consistent with expectations based on observations over the last several solar cycles. Cycles~4 and 19 are notable exceptions, exhibiting a significant lag in the field reversal. Interestingly, these are among the strongest cycles observed. In particular, Cycle~19 stands out as the strongest and is followed by a significantly weaker cycle. Strong cycles begin with strong polar fields. This requires a significant amount of flux to be transported to the poles in order to cancel the existing field and buildup the opposite-polarity field for the next cycle. This effect results in a weaker polar field at the end of the cycle, leading to a weaker following cycle. This may suggest that there is a fundamental limit to how much the polar fields can change from one cycle to the next, representing a saturation method that prevents solar cycles from becoming too large.

Finally, in \autoref{fig:tsi_all}b, we also note multiple surges of opposite-polarity streams going towards the poles in the magnetic butterfly diagram, which we also often see in observed data \citep{Upton2014, Upton2014a, Jha2024a} and in reconstructed butterfly diagrams based on proxies of magnetic field \citep[Ca II K and H-alpha filaments;][]{Mordvinov2022}. A similar plot showing the variation of the polar field and ADM for $\Delta\gamma = 0$ is presented in \autoref{fig:tsi_all_app}(a) and \ref{fig:tsi_all_app}(b).

We emphasize that the present findings are based on a single SARG realization with a modified tilt, chosen to bring the ADM into agreement with the expected ADM at the end of the cycle. In our follow-up work, we plan to address uncertainty quantification through a global sensitivity analysis of AFT, using ensembles of SARG realizations and varying additional parameters beyond tilt—such as cycle asymmetry, the presence of anti-Hale regions, and others.

\subsection{Comparison with Recent Observations}

In this work, we use the mean of the tilt distribution as a tuning parameters to fix the ADM at the end of each cycle. For a consistency check, we calculate the polar field strengths in our reconstructed photospheric magnetic field maps ($B_r$) and compare it to the polar field during the modern observational era, either directly or through established proxies. These include polar faculae counts \citep{Munoz2012} and the Polar Network Index \citep[PNI;][]{Mishra2025}, both widely used as proxies for the polar magnetic field strength.

\begin{figure}[!htbp]
    \centering
    \includegraphics[width=0.45\textwidth]{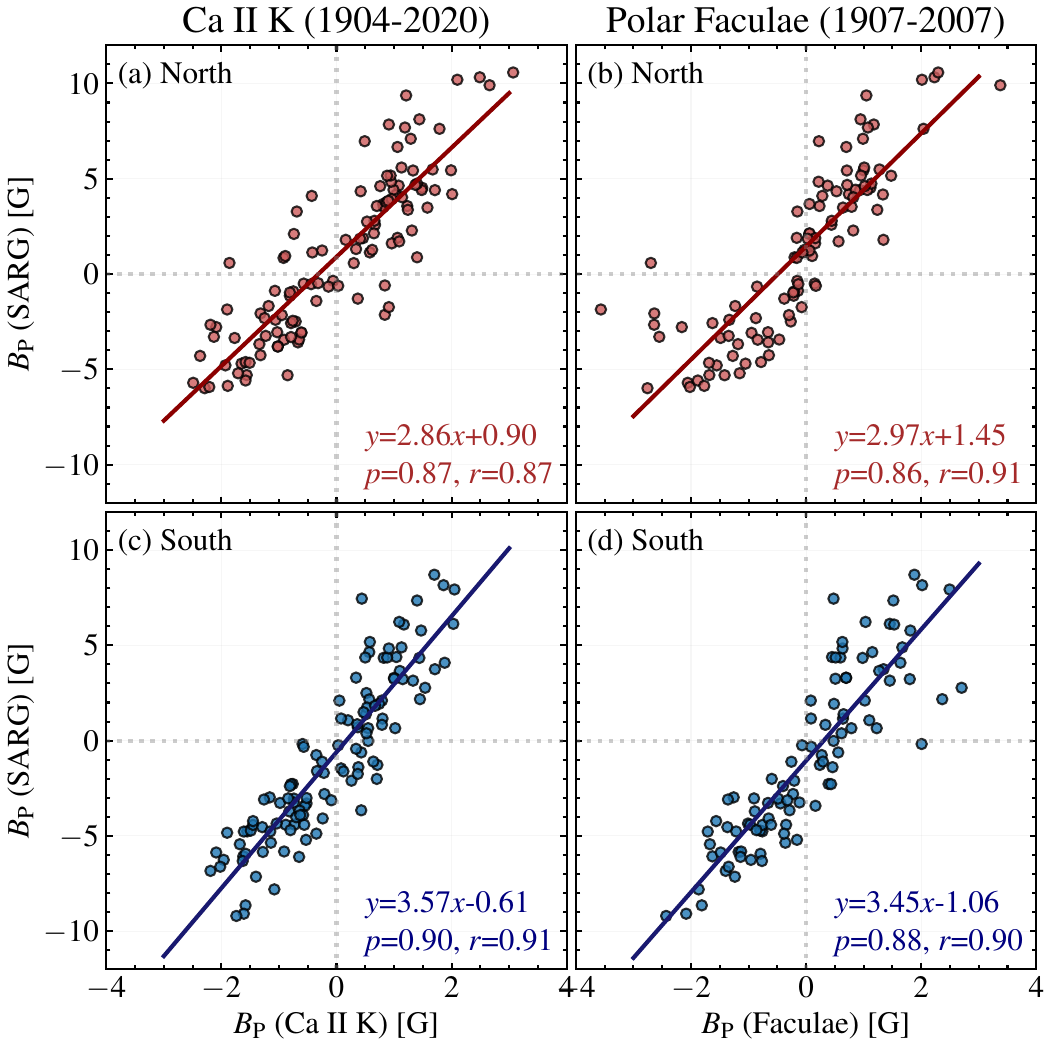}
    \caption{The scatter plot showing the correlation between the polar field reconstructed (with non-zero $\Delta \gamma$) here and its correlation with the other proxies of polar field obtained from KoSO Ca II K and faculae counts in the northern hemisphere (a, b) and for southern hemisphere (c, d) .}
    \label{fig:polar}
\end{figure}

In \autoref{fig:polar}(a\,--\,d), we present correlation plots comparing our derived polar field with polar faculae counts \bkj{and PNI} over their overlapping observational period. Additionally, in \autoref{tab:correlation} we show the correlations in the original case (where we don't modify the average $\gamma$ in tilts) and for the case where we shift the $\gamma$ to get the desired ADM. We also show the correlations for each proxy in the northern and southern hemispheres separately. These results show excellent agreement, with Pearson correlation coefficients $(p = 0.86)$ and Spearman`s correlation coefficient $(r = 0.91)$ in the northern hemisphere and $(p = 0.88, r = 0.9)$ in the southern hemisphere. Similarly, when comparing our reconstructed polar field with the PNI-based estimates, we obtain nearly identical correlation coefficients: $(p = 0.87, r = 0.87)$ in the north and \(p = 0.90, r = 0.90\) in the south. These high correlation values strongly support the fidelity of the SARG-generated active region catalog with observational proxies, at least at global (polar) scales. However, in all cases (\autoref{fig:polar}(a\,--\,d)), we note that the slope of the best-fit line ranges from 2.86\,--\,3.57, indicating that the polar field values derived in this work are systematically scaled higher compared to those inferred from the proxy datasets. We also observe a consistently higher scaling factor in the southern hemisphere in both datasets.

Fixing the ADM to match with the observation does not ensure that the polar fields will also match the observations. This arises from the fact that there are many possible polar field configurations that can produce a specific ADM value. While, factors such as meridional flow speeds, flow and tilt quenching function, separation distances between the two polarities of AR, choice of the morphology of AR and the distribution of magnetic field all contribute to the cancellation and buildup of the polar field. Any one of these parameters can be tuned to achieve the desired ADM or modulate the polar field. For example, the morphology of ARs plays a vital role in the flux cancellation process, which in our case is an idealized 2D Gaussian instead of the complex structure observed on the Sun. Changing the morphology, or even just the distance between the bipoles will change the amount of flux that reaches the poles. In our case, we chose to modulate the ADM with mean tilt. While we could alter any of these parameters to reduce the strength of polar fields, this would also reduce the amplitude of ADM, countering our efforts to fix the ADM.

%


\begin{table}[htbp!]
    \centering
    \caption{Pearson Correlation coefficients ($p$) and  Spearman`s correlation coefficient ($r$) calculated in the polar field derived in this work against the faculae count and Ca II K PNI in northern and southern hemispheres, for $\Delta\gamma = 0$ and $\Delta\gamma \ne 0$.}
    \label{tab:correlation}
    \begin{tabular}{|c|cc|cc|cc|cc|}
        \hline
        & \multicolumn{4}{c|}{Faculae (1907\,--\,2007)\tablenote{\citet{Munoz2012}}} & \multicolumn{4}{c|}{Ca II K (1904\,--\,2020)\tablenote{\citet{Mishra2025}}} \\
        \cline{2-9}
        $\Delta\gamma$  & \multicolumn{2}{c|}{North} & \multicolumn{2}{c|}{South} & \multicolumn{2}{c|}{North} & \multicolumn{2}{c|}{South} \\
        \cline{2-9}
        & $p$ & $r$ & $p$ & $r$ & $p$ & $r$ & $p$ & $r$ \\
        \hline
        $=0$   & 0.80 & 0.89 & 0.87 & 0.88 & 0.82 & 0.84 & 0.89 & 0.89 \\
        $\ne 0$ & 0.86 & 0.91 & 0.88 & 0.90 & 0.87 & 0.87 & 0.90 & 0.91 \\
        \hline
    \end{tabular}
\end{table}

\begin{figure*}[htbp!]
    \centering
    \includegraphics[width=\textwidth]{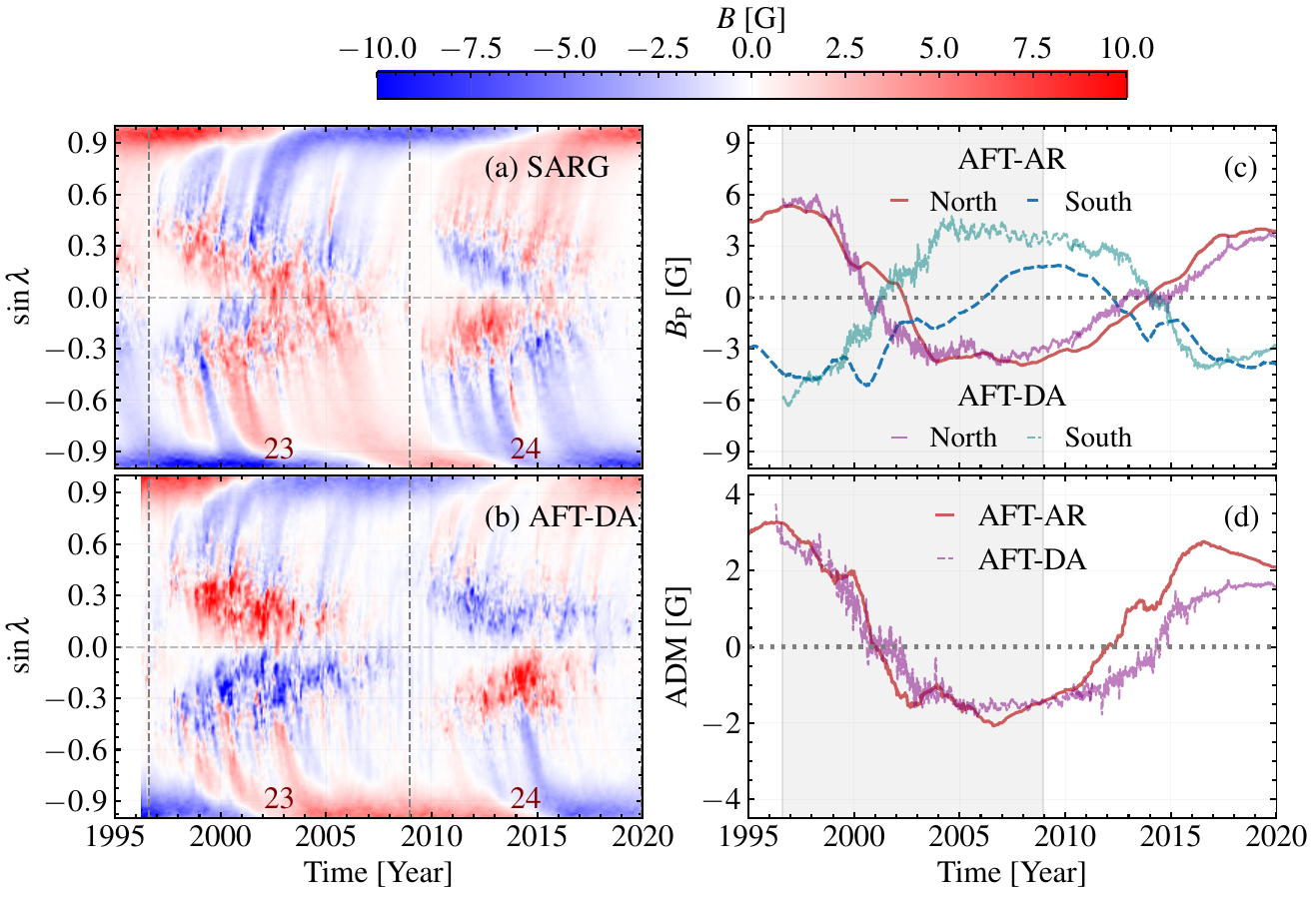}
    \caption{Comparison between the AFT-ARSynth simulation (bold lines) and the AFT-DA (MDI and HMI) simulation (shaded lines). Panels show the magnetic butterfly diagrams constructed by azimuthally averaging the radial magnetic field over each Carrington rotation for: (a) AFT-ARSynth using the SARG catalog, and (b) AFT-DA from 1996\,--\,2020. Panel (c) shows the comparison of the polar field, computed above $60^\circ$ latitude in each hemisphere, while panel (d) presents the ADM for both simulation modes. solid and dashed in this plot}
    \label{fig:aftcomp}
\end{figure*}

To further validate our historical reconstruction, we make a more detailed comparison of our synthetic simulation against the AFT Baseline (which has been shown to closely match HMI observations). In \autoref{fig:aftcomp}a and \ref{fig:aftcomp}b, we show the magnetic butterfly diagrams generated using AFT-ARSynth and AFT-DA (MDI and HMI), respectively, and compare them over their overlapping time range (1996\,--\,2020). One of the most notable qualitative features is the hemispheric asymmetry, an intrinsic characteristic of the solar cycle, that appears in the data-driven model but is absent in the synthetic reconstruction. This is expected, as hemispheric asymmetry was not explicitly incorporated into the SARG model. This is expected, as hemispheric asymmetry was not explicitly incorporated into the SARG model. However, we note that polar streams of the strong magnetic flux do not always show a one-to-one correspondence.  These polar streams cane be driven by the emergence of strong or highly tilted AR. Due to intentionally added randomness in SARG, we don't expect to produce ARs with exactly the same set of characteristics as the observed ARs and hence the difference in the strength and timing of these streams is not surprising. However, we note that another mechanism for achieving these streams is brief gaps in AR emergence. Observed similarities in some of the streams is evidence of this.

While the polar fields strengths are in relatively good qualitative agreement for most of the time,  particularly in the North, there are some  notable discrepancies. For instance, the polar field in the South (blue) deviates significantly from the observation (AFT-DA) from 200-2015. This produces a significant lag in the timing of polar field reversals during Cycle~23, where the south pole reverses several years after north(\autoref{fig:aftcomp}c). In contrast, Cycle~24 does not exhibit such a pronounced discrepancy as both hemispheres reverse polarity at nearly the same time, albeit a couple of years prior to the observed reversal. Despite these differences, the comparison of the ADM from these two simulation, shown in \autoref{fig:aftcomp}d, reveals a fairly good agreement between the two simulations for most of these two cycles. We do see that the ADM begins to go astray near the end, but then heads back into agreement. This is somewhat expected as ADM is used as our measuring scale in this reconstruction. Given the inherent complexity and stochastic nature of the processes governing AR emergence, especially the role of tilt angles in polar field evolution, we do not expect a one-to-one match between the two simulations. In this case a significant source of the discrepancies may be the notable hemispheric asymmetry during Cycle~24. Currently, SARG uses the global SSN, but incorporating the hemispheric differences will likely produce even better agreement. We leave this for a future investigation. Nevertheless, the results suggest that our synthetic reconstruction captures the large-scale behavior fairly well considering it was created entirely with synthetic ARs prescribed by SARG with only the global SSN.


\section{Summary}

The reconstruction of historical solar magnetic fields is important for a wide range of studies that require not only global solar properties but also full photospheric magnetic field maps ($B_r$). These maps enable long-term investigations of solar variability over multiple centuries and can serve as input for coronal and heliospheric solar wind models to study the extended evolution of the solar atmosphere. To this end, we employed our newly developed Synthetic Active Region Generator (SARG) in combination with the improved \ac{aft2} model to simulate $B_r$ from Cycle~1 through Cycle~24. The reconstructed maps show excellent agreement with the expected features of solar cycle variability, including polarity reversal near solar maximum due to the cancellation of existing polarity and the buildup of opposite-polarity flux in the polar regions through the poleward migration of diffused magnetic field. Additionally, we observe multiple surges of magnetic flux that lead to local fluctuations in the polar field during several cycles.

We compared the polar field estimates from our reconstruction with two widely used proxies: polar faculae counts and the Ca II K PNI. Since both of these proxies are calibrated against WSO polar field measurements, good correlations are expected. Our results confirm this, showing correlation coefficients exceeding $0.85$ in both hemispheres for each proxy. These high correlations provide strong validation for our synthetic active region approach.

The comparison with the data-driven AFT-DA simulation highlights both the fidelity and the limitations of our synthetic reconstruction approach. While the SARG-based historical reconstruction successfully captures large-scale features of the solar magnetic field evolution, it is inherently challenging to reproduce all observed quantities with high accuracy due to the complex and nonlinear nature of the underlying physical processes, as well as the intrinsic variability within the statistical relationships employed in the model. 

For instance, \citet{DasiEspuig2010} demonstrated an anti-correlation between the mean tilt angle of active regions and the strength of the solar cycle. Additionally, recent studies \citep{Jha2020, Karak2020,Sreedevi2023, sreedevi_analysis_2024} have shown that the tilt of active regions depends on their magnetic field strength and flux, with evidence of saturation effects influenced by latitude. Although SARG incorporates a cyclic dependence of the emergence latitude \citep{Hathaway2011a}, other important dependencies remain to be modeled. In future work, we plan to further refine and validate the SARG-based AFT-AR.SYNTH model by incorporating these physical dependencies. This will enable a deeper understanding of the complexity and nonlinearity in solar magnetic field evolution as captured through flux transport simulations. Thus, in future, we plan to make use of the all the existing sunspot catalogs \citep{Balmaceda2009, Mandal2017, Mandal2020, Jha2022} compiled from various sources along with incorporating these characteristics reported in various studies  to develop an advance of SARG with more data driven approach.

The historical photospheric magnetic field maps simulated using \ac{aft2} will be made publicly available and can be also obtained by sending a request to corresponding author. Hundreds of SARG realizations is publicly released under a Creative Commons license via Zenodo \citep[][\url{https://zenodo.org/records/16915548}]{jha2025_sarg}. Furthermore, a custom SARG realizations with additional constraints can be generated upon request by directly contacting the corresponding author.

\begin{acknowledgments}
BKJ and LAU were supported by NASA Heliophysics Living With a Star Strategic Capabilites grant NNH21ZDA001N-LWSSC and NASA grant NNH18ZDA001N-DRIVE to the COFFIES DRIVE Center managed by Stanford University. L.A.U. was supported by NASA Heliophysics Guest Investigator NNH15ZDA001N-HGI and Solar Irradiance Science Team grant 80NSSC18K1501. GK gratefully acknowledges support of the NASA Solar Irradiance Science Team program via grant 80NSSC18K1501. This research has made use of NASA's Astrophysics Data System (ADS; \url{https://ui.adsabs.harvard. edu/}) Bibliographic Services. We also thank the anonymous referee for their valuable comments and suggestions, which helped improve the quality of this work.

\end{acknowledgments}


\begin{contribution}

LAU developed the SARG model. BKJ and LAU jointly developed the AFT~2.0 model (on top of AFT developed by LAU), performed the full set of simulations, and analyzed the data used in this study. BKJ led the manuscript preparation, with support from LAU in writing and revising the text. GC and OC provided critical input for the simulations and contributed to the drafting of the manuscript.
\end{contribution}

\software{Matplotlib \citep{Hunter2007}, Numpy \citep{harris2020array} and Pandas \citep{pandas2020}}

\appendix
\section{SARG: Butterfly Diagram}
\label{app:butterfly}

\begin{sidewaysfigure*}[htbp!]
    \includegraphics[width=\textheight]{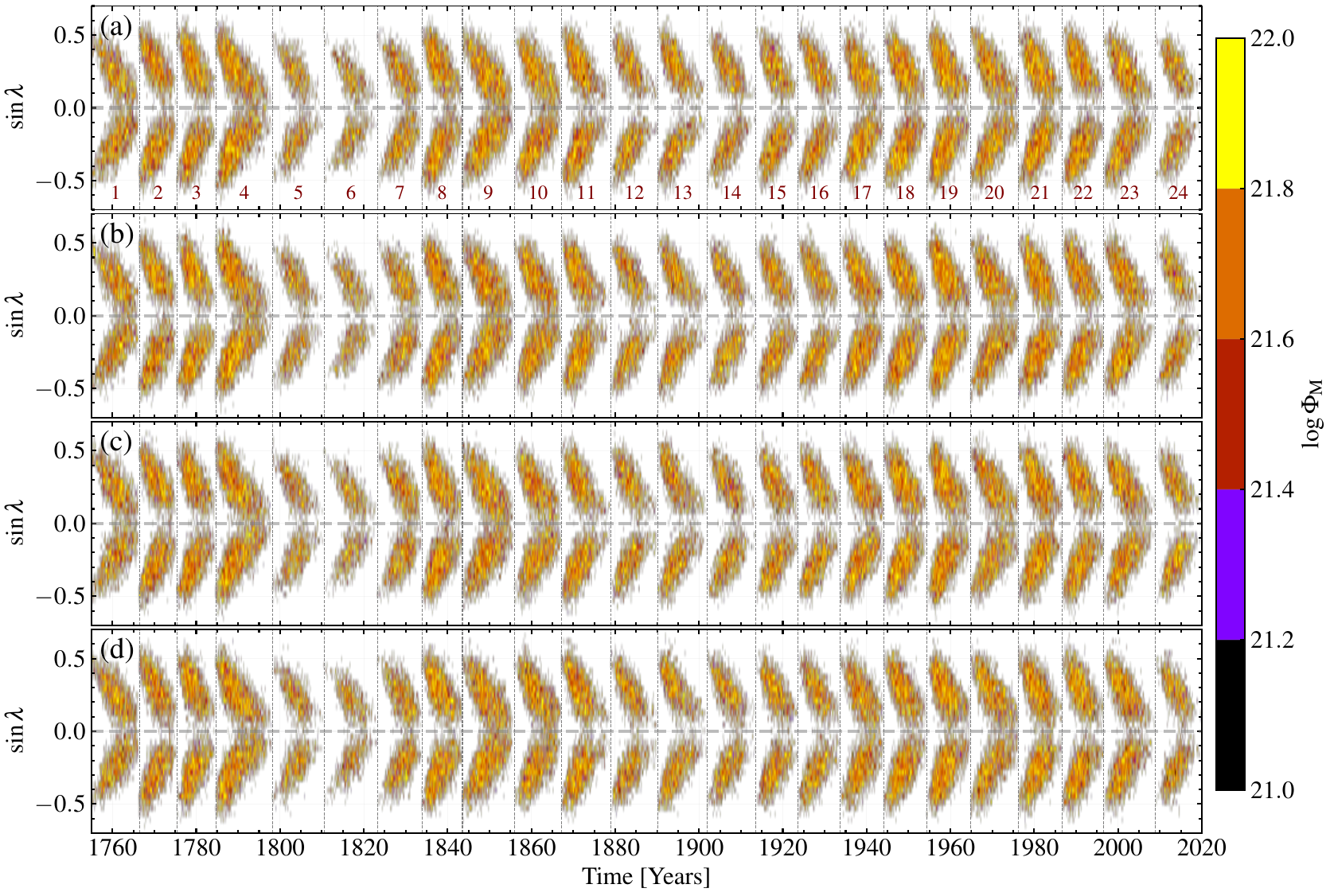}
    \caption{ AR butterfly diagrams constructed by averaging the magnetic flux in fifty uniform $\sin{\lambda}$ bins and further averaging over each Carrington rotation (27.3 days). The four panels show butterfly diagrams generated in the same manner for four randomly selected realizations of the SARG model, illustrating both the statistical similarity and small-scale differences among them.}
    \label{fig:sarg_butter}
\end{sidewaysfigure*}

Each realization of the SARG model produces a distinct active region (AR) catalog due to the inclusion of stochastic elements at multiple stages of the algorithm. While individual realizations differ, they all adhere to the same underlying statistical laws. In \autoref{fig:sarg_butter}, we present the magnetic butterfly diagrams for four different SARG realizations. These diagrams are constructed by computing the mean magnetic flux in 50 uniformly spaced $\sin{\lambda}$ (sine of latitude) bins, averaged over each Carrington rotation (27.3\,days). The similarities and differences among the panels illustrate the statistical consistency across realizations, while the small-scale variations reflect granular differences introduced by the random processes in the SARG algorithm. 

\section{Modification of tilt in a SARG Catalog}
\label{app:tiltmod}

In a SARG catalog, tilt modification is implemented by adjusting the tilt angle of each AR by a specified amount. For a desired tilt modification of $\Delta\gamma$, the original tilt $\gamma$ is modified to $\gamma^\prime$ according to the following hemispheric-dependent relation:
\begin{equation}
\gamma^\prime =
\begin{cases}
    \gamma + \Delta\gamma, & \text{if}~\lambda \ge 0, \\
    \gamma - \Delta\gamma, & \text{if}~\lambda < 0,
\end{cases}
\end{equation}
where $\lambda$ is the latitude of emergence.

\begin{figure}[htbp!]
    \centering
    \includegraphics[width=0.9\textwidth]{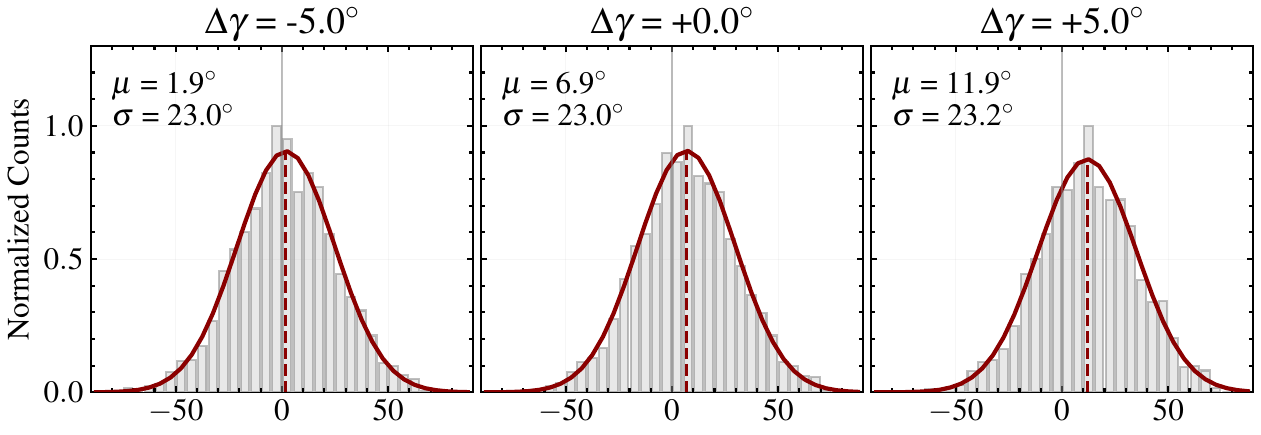}
    \caption{Distribution of active region tilt angles for Solar Cycle~10, shown for cases where the tilt of each AR is modified by the value indicated in each panel, ranging from $-5^\circ$ (left), actual $0^\circ$ (middle) $+5^\circ$ (right). The dark solid curve in each panel represents the best-fit Gaussian distribution, with the corresponding mean and standard deviation ($\sigma$) labeled. The solid and dashed lines represent the 0-axis line and mean of the distribution as calculated from Gaussian best fit.}
    \label{fig:sarg_modtilt}
\end{figure}

The effects of varying $\Delta\gamma$ are illustrated in \autoref{fig:sarg_modtilt}. For each case, the tilt angles of individual ARs in the catalog are modified using the above relation, and the positions of leading and following polarities are recalculated following Step~6 in \autoref{sec:sarg}. From \autoref{fig:sarg_modtilt}, we observe that modifying the tilt systematically shifts the mean of the tilt angle distribution, as determined by fitting a Gaussian function. The shift in the mean closely matches the applied $\Delta\gamma$, while the standard deviation ($\sigma$) of the distribution remains largely unchanged. This demonstrates that the modification alters the central tendency without affecting the statistical spread of the tilt angles.

\begin{sidewaysfigure*}[htbp!]
    \includegraphics[width=0.96\textheight]{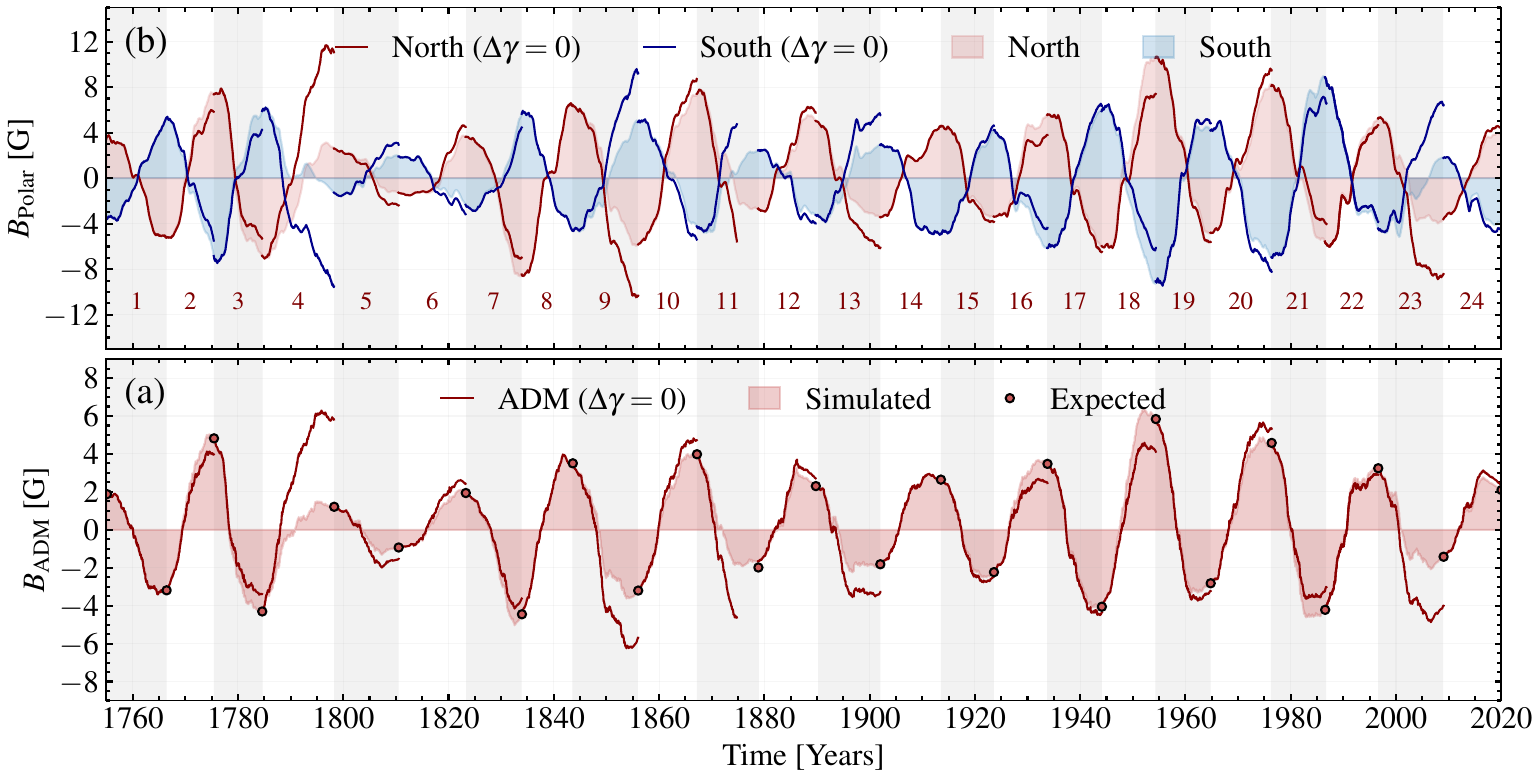}
    \caption{Similar to \autoref{fig:tsi_all}, the case of $\Delta\gamma = 0$ and $\Delta\gamma \ne 0$. Here each cycle starts with the initial condition that matches the expected ADM at the end of the previous cycle. (b) The polar field calculated over latitude 60$^\circ$ and (d) the simulated ADM values along with that expected from the empirical relation given in \autoref{fig:admssnl}.}
    \label{fig:tsi_all_app}
\end{sidewaysfigure*}

In \autoref{fig:tsi_all_app} we show a plot similar to the \autoref{fig:tsi_all} for the case of $\Delta\gamma = 0$, where at the end of each cycle we replace the initial condition with the expected ADM with $\Delta\gamma \ne 0$.

\bibliography{formated_all}{}

\begin{thebibliography}{}
\expandafter\ifx\csname natexlab\endcsname\relax\def\natexlab#1{#1}\fi
\providecommand{\url}[1]{\href{#1}{#1}}
\providecommand{\dodoi}[1]{doi:~\href{http://doi.org/#1}{\nolinkurl{#1}}}
\providecommand{\doeprint}[1]{\href{http://ascl.net/#1}{\nolinkurl{http://ascl.net/#1}}}
\providecommand{\doarXiv}[1]{\href{https://arxiv.org/abs/#1}{\nolinkurl{https://arxiv.org/abs/#1}}}

\bibitem[{H.~W. {Babcock}(1953){Babcock}}]{Babcock1953}
{Babcock}, H.~W. 1953, \bibinfo{title}{{The Solar Magnetograph.},} \apj, 118, 387, \dodoi{10.1086/145767}

\bibitem[{L.~A. {Balmaceda} {et~al.}(2009){Balmaceda}, {Solanki}, {Krivova}, \& {Foster}}]{Balmaceda2009}
{Balmaceda}, L.~A., {Solanki}, S.~K., {Krivova}, N.~A., \& {Foster}, S. 2009, \bibinfo{title}{{A homogeneous database of sunspot areas covering more than 130 years},} Journal of Geophysical Research (Space Physics), 114, A07104, \dodoi{10.1029/2009JA014299}

\bibitem[{R.~M. {Caplan} {et~al.}(2025){Caplan}, {Stulajter}, {Linker}, {Downs}, {Upton}, {Jha}, {Attie}, {Arge}, \& {Henney}}]{Caplan2025}
{Caplan}, R.~M., {Stulajter}, M.~M., {Linker}, J.~A., {et~al.} 2025, \bibinfo{title}{{Open-source Flux Transport (OFT). I. HipFT{\textendash}High-performance Flux Transport},} \apjs, 278, 24, \dodoi{10.3847/1538-4365/adc080}

\bibitem[{P. {Charbonneau}(2010){Charbonneau}}]{Charbonneau2010}
{Charbonneau}, P. 2010, \bibinfo{title}{{Dynamo Models of the Solar Cycle},} Liv. Rev. Sol. Phys., 7, 3, \dodoi{10.12942/lrsp-2010-3}

\bibitem[{P. Charbonneau(2014)Charbonneau}]{Charbonneau2014}
Charbonneau, P. 2014, \bibinfo{title}{Solar Dynamo Theory,} Ann. Rev. Astron. Astrophys., 52, 251, \dodoi{10.1146/annurev-astro-081913-040012}

\bibitem[{F. {Clette} {et~al.}(2016){Clette}, {Cliver}, {Lef{\`e}vre}, {Svalgaard}, {Vaquero}, \& {Leibacher}}]{Clette2016}
{Clette}, F., {Cliver}, E.~W., {Lef{\`e}vre}, L., {et~al.} 2016, \bibinfo{title}{{Preface to Topical Issue: Recalibration of the Sunspot Number},} \solphys, 291, 2479, \dodoi{10.1007/s11207-016-1017-8}

\bibitem[{F. {Clette} \& L. {Lefèvre}(2015){Clette} \& {Lefèvre}}]{ssn}
{Clette}, F., \& {Lefèvre}, L. 2015, \bibinfo{title}{SILSO Sunspot Number V2.0,}, https://doi.org/10.24414/qnza-ac80

\bibitem[{M. {Dasi-Espuig} {et~al.}(2010){Dasi-Espuig}, {Solanki}, {Krivova}, {Cameron}, \& {Pe{\~n}uela}}]{DasiEspuig2010}
{Dasi-Espuig}, M., {Solanki}, S.~K., {Krivova}, N.~A., {Cameron}, R., \& {Pe{\~n}uela}, T. 2010, \bibinfo{title}{{Sunspot group tilt angles and the strength of the solar cycle},} \aap, 518, A7, \dodoi{10.1051/0004-6361/201014301}

\bibitem[{L.~J. {Gray} {et~al.}(2010){Gray}, {Beer}, {Geller}, {Haigh}, {Lockwood}, {Matthes}, {Cubasch}, {Fleitmann}, {Harrison}, {Hood}, {Luterbacher}, {Meehl}, {Shindell}, {van Geel}, \& {White}}]{Gray2010}
{Gray}, L.~J., {Beer}, J., {Geller}, M., {et~al.} 2010, \bibinfo{title}{{Solar Influences on Climate},} Reviews of Geophysics, 48, RG4001, \dodoi{10.1029/2009RG000282}

\bibitem[{G.~E. {Hale}(1908){Hale}}]{Hale1908}
{Hale}, G.~E. 1908, \bibinfo{title}{{On the Probable Existence of a Magnetic Field in Sun-Spots},} \apj, 28, 315, \dodoi{10.1086/141602}

\bibitem[{G.~E. {Hale} {et~al.}(1919){Hale}, {Ellerman}, {Nicholson}, \& {Joy}}]{Hale1919}
{Hale}, G.~E., {Ellerman}, F., {Nicholson}, S.~B., \& {Joy}, A.~H. 1919, \bibinfo{title}{{The Magnetic Polarity of Sun-Spots},} \apj, 49, 153, \dodoi{10.1086/142452}

\bibitem[{C.~R. Harris {et~al.}(2020)Harris, Millman, van~der Walt, Gommers, Virtanen, Cournapeau, Wieser, Taylor, Berg, Smith, Kern, Picus, Hoyer, van Kerkwijk, Brett, Haldane, del R{\'{i}}o, Wiebe, Peterson, G{\'{e}}rard-Marchant, Sheppard, Reddy, Weckesser, Abbasi, Gohlke, \& Oliphant}]{harris2020array}
Harris, C.~R., Millman, K.~J., van~der Walt, S.~J., {et~al.} 2020, \bibinfo{title}{Array programming with {NumPy},} Nature, 585, 357, \dodoi{10.1038/s41586-020-2649-2}

\bibitem[{D. Hathaway {et~al.}(2025)Hathaway, Upton, \& Jha}]{hathaway_2025_17108109}
Hathaway, D., Upton, L., \& Jha, B.~K. 2025, \bibinfo{title}{A Unified Sunspot Group Dataset (RGO/SOON/NOAA ) from 1874 to Present,}, 1.0.0 Zenodo, \dodoi{10.5281/zenodo.17108109}

\bibitem[{D.~H. {Hathaway}(2011){Hathaway}}]{Hathaway2011a}
{Hathaway}, D.~H. 2011, \bibinfo{title}{{A Standard Law for the Equatorward Drift of the Sunspot Zones},} \solphys, 273, 221, \dodoi{10.1007/s11207-011-9837-z}

\bibitem[{D.~H. {Hathaway}(2015){Hathaway}}]{Hathaway2015}
{Hathaway}, D.~H. 2015, \bibinfo{title}{{The Solar Cycle},} Living Reviews in Solar Physics, 12, 4, \dodoi{10.1007/lrsp-2015-4}

\bibitem[{D.~H. {Hathaway} \& L. {Rightmire}(2011){Hathaway} \& {Rightmire}}]{Hathaway2011}
{Hathaway}, D.~H., \& {Rightmire}, L. 2011, \bibinfo{title}{{Variations in the Axisymmetric Transport of Magnetic Elements on the Sun: 1996-2010},} \apj, 729, 80, \dodoi{10.1088/0004-637X/729/2/80}

\bibitem[{D.~H. {Hathaway} \& L.~A. {Upton}(2016){Hathaway} \& {Upton}}]{Hathaway2016}
{Hathaway}, D.~H., \& {Upton}, L.~A. 2016, \bibinfo{title}{{Predicting the amplitude and hemispheric asymmetry of solar cycle 25 with surface flux transport},} Journal of Geophysical Research (Space Physics), 121, 10,744, \dodoi{10.1002/2016JA023190}

\bibitem[{K.~S. {Hickmann} {et~al.}(2015){Hickmann}, {Godinez}, {Henney}, \& {Arge}}]{hickmann2015}
{Hickmann}, K.~S., {Godinez}, H.~C., {Henney}, C.~J., \& {Arge}, C.~N. 2015, \bibinfo{title}{{Data Assimilation in the ADAPT Photospheric Flux Transport Model},} \solphys, 290, 1105, \dodoi{10.1007/s11207-015-0666-3}

\bibitem[{R. {Howard}(1974){Howard}}]{Howard1974}
{Howard}, R. 1974, \bibinfo{title}{{Studies of Solar Magnetic Fields. I: The Average Field Strengths},} \solphys, 38, 283, \dodoi{10.1007/BF00155067}

\bibitem[{J.~D. Hunter(2007)Hunter}]{Hunter2007}
Hunter, J.~D. 2007, \bibinfo{title}{Matplotlib: A 2D graphics environment,} Computing in Science \& Engineering, 9, 90, \dodoi{10.1109/MCSE.2007.55}

\bibitem[{B.~K. {Jha} {et~al.}(2022){Jha}, {Hegde}, {Priyadarshi}, {Mandal}, {Ravindra}, \& {Banerjee}}]{Jha2022}
{Jha}, B.~K., {Hegde}, M., {Priyadarshi}, A., {et~al.} 2022, \bibinfo{title}{{Extending the sunspot area series from Kodaikanal Solar Observatory},} Frontiers in Astronomy and Space Sciences, 9, 1019751, \dodoi{10.3389/fspas.2022.1019751}

\bibitem[{B.~K. {Jha} {et~al.}(2020){Jha}, {Karak}, {Mandal}, \& {Banerjee}}]{Jha2020}
{Jha}, B.~K., {Karak}, B.~B., {Mandal}, S., \& {Banerjee}, D. 2020, \bibinfo{title}{{Magnetic Field Dependence of Bipolar Magnetic Region Tilts on the Sun: Indication of Tilt Quenching},} \apjl, 889, L19, \dodoi{10.3847/2041-8213/ab665c}

\bibitem[{B.~K. Jha \& L. Upton(2025)Jha \& Upton}]{jha2025_sarg}
Jha, B.~K., \& Upton, L. 2025, \bibinfo{title}{Solar Active Region Emergence Catalog from the Synthetic Active Region Generator (SARG),}, 1.0.0 Zenodo, \dodoi{10.5281/zenodo.16915548}

\bibitem[{B.~K. {Jha} \& L.~A. {Upton}(2024){Jha} \& {Upton}}]{Jha2024a}
{Jha}, B.~K., \& {Upton}, L.~A. 2024, \bibinfo{title}{{Predicting the Timing of the Solar Cycle 25 Polar Field Reversal},} \apjl, 962, L15, \dodoi{10.3847/2041-8213/ad20d2}

\bibitem[{J. Jiang {et~al.}(2011{\natexlab{a}})Jiang, Cameron, Schmitt, \& Sch{\"u}ssler}]{Jiang2011}
Jiang, J., Cameron, R.~H., Schmitt, D., \& Sch{\"u}ssler, M. 2011{\natexlab{a}}, \bibinfo{title}{The Solar Magnetic Field since 1700 - {{I}}. {{Characteristics}} of Sunspot Group Emergence and Reconstruction of the Butterfly Diagram,} Astronomy \& Astrophysics, 528, A82, \dodoi{10.1051/0004-6361/201016167}

\bibitem[{J. Jiang {et~al.}(2011{\natexlab{b}})Jiang, Cameron, Schmitt, \& Sch{\"u}ssler}]{Jiang2011a}
Jiang, J., Cameron, R.~H., Schmitt, D., \& Sch{\"u}ssler, M. 2011{\natexlab{b}}, \bibinfo{title}{The Solar Magnetic Field since 1700 - {{I}}. {{Characteristics}} of Sunspot Group Emergence and Reconstruction of the Butterfly Diagram,} Astronomy \& Astrophysics, 528, A82, \dodoi{10.1051/0004-6361/201016167}

\bibitem[{J. {Jiang} {et~al.}(2014){Jiang}, {Hathaway}, {Cameron}, {Solanki}, {Gizon}, \& {Upton}}]{Jiang2014a}
{Jiang}, J., {Hathaway}, D.~H., {Cameron}, R.~H., {et~al.} 2014, \bibinfo{title}{{Magnetic Flux Transport at the Solar Surface},} \ssr, 186, 491, \dodoi{10.1007/s11214-014-0083-1}

\bibitem[{B.~B. {Karak}(2020){Karak}}]{Karak2020}
{Karak}, B.~B. 2020, \bibinfo{title}{{Dynamo Saturation through the Latitudinal Variation of Bipolar Magnetic Regions in the Sun},} \apjl, 901, L35, \dodoi{10.3847/2041-8213/abb93f}

\bibitem[{B.~B. {Karak}(2023){Karak}}]{Karak2023}
{Karak}, B.~B. 2023, \bibinfo{title}{{Models for the long-term variations of solar activity},} Living Reviews in Solar Physics, 20, 3, \dodoi{10.1007/s41116-023-00037-y}

\bibitem[{G. {Kopp} {et~al.}(2016){Kopp}, {Krivova}, {Wu}, \& {Lean}}]{Greg2016}
{Kopp}, G., {Krivova}, N., {Wu}, C.~J., \& {Lean}, J. 2016, \bibinfo{title}{{The Impact of the Revised Sunspot Record on Solar Irradiance Reconstructions},} \solphys, 291, 2951, \dodoi{10.1007/s11207-016-0853-x}

\bibitem[{W.~C. {Livingston} {et~al.}(1976){Livingston}, {Harvey}, {Slaughter}, \& {Trumbo}}]{Livingston1976}
{Livingston}, W.~C., {Harvey}, J., {Slaughter}, C., \& {Trumbo}, D. 1976, \bibinfo{title}{{Solar magnetograph employing integrated diode arrays.},} \ao, 15, 40, \dodoi{10.1364/AO.15.000040}

\bibitem[{S. {Mandal} {et~al.}(2017){Mandal}, {Karak}, \& {Banerjee}}]{Mandal2017}
{Mandal}, S., {Karak}, B.~B., \& {Banerjee}, D. 2017, \bibinfo{title}{{Latitude Distribution of Sunspots: Analysis Using Sunspot Data and a Dynamo Model},} \apj, 851, 70, \dodoi{10.3847/1538-4357/aa97dc}

\bibitem[{S. {Mandal} {et~al.}(2020){Mandal}, {Krivova}, {Solanki}, {Sinha}, \& {Banerjee}}]{Mandal2020}
{Mandal}, S., {Krivova}, N.~A., {Solanki}, S.~K., {Sinha}, N., \& {Banerjee}, D. 2020, \bibinfo{title}{{Sunspot area catalog revisited: Daily cross-calibrated areas since 1874},} \aap, 640, A78, \dodoi{10.1051/0004-6361/202037547}

\bibitem[{D.~K. {Mishra} {et~al.}(2025){Mishra}, {Jha}, {Chatzistergos}, {Ermolli}, {Banerjee}, {Upton}, \& {Khan}}]{Mishra2025}
{Mishra}, D.~K., {Jha}, B.~K., {Chatzistergos}, T., {et~al.} 2025, \bibinfo{title}{{Ca II K Polar Network Index of the Sun: A Proxy for Historical Polar Magnetic Field},} \apj, 982, 78, \dodoi{10.3847/1538-4357/adb3a8}

\bibitem[{A.~V. {Mordvinov} {et~al.}(2022){Mordvinov}, {Karak}, {Banerjee}, {Golubeva}, {Khlystova}, {Zhukova}, \& {Kumar}}]{Mordvinov2022}
{Mordvinov}, A.~V., {Karak}, B.~B., {Banerjee}, D., {et~al.} 2022, \bibinfo{title}{{Evolution of the Sun's activity and the poleward transport of remnant magnetic flux in Cycles 21-24},} \mnras, 510, 1331, \dodoi{10.1093/mnras/stab3528}

\bibitem[{J.~M. {Mosher}(1977){Mosher}}]{Mosher1977}
{Mosher}, J.~M. 1977, PhD thesis, California Institute of Technology

\bibitem[{A. {Mu{\~n}oz-Jaramillo} {et~al.}(2021){Mu{\~n}oz-Jaramillo}, {Navarrete}, \& {Campusano}}]{MunozJaramillo2021}
{Mu{\~n}oz-Jaramillo}, A., {Navarrete}, B., \& {Campusano}, L.~E. 2021, \bibinfo{title}{{Solar Anti-Hale Bipolar Magnetic Regions: A Distinct Population with Systematic Properties},} \apj, 920, 31, \dodoi{10.3847/1538-4357/ac133b}

\bibitem[{A. {Mu{\~n}oz-Jaramillo} {et~al.}(2012){Mu{\~n}oz-Jaramillo}, {Sheeley}, {Zhang}, \& {DeLuca}}]{Munoz2012}
{Mu{\~n}oz-Jaramillo}, A., {Sheeley}, N.~R., {Zhang}, J., \& {DeLuca}, E.~E. 2012, \bibinfo{title}{{Calibrating 100 Years of Polar Faculae Measurements: Implications for the Evolution of the Heliospheric Magnetic Field},} \apj, 753, 146, \dodoi{10.1088/0004-637X/753/2/146}

\bibitem[{A. {Mu{\~n}oz-Jaramillo} {et~al.}(2015){Mu{\~n}oz-Jaramillo}, {Senkpeil}, {Windmueller}, {Amouzou}, {Longcope}, {Tlatov}, {Nagovitsyn}, {Pevtsov}, {Chapman}, {Cookson}, {Yeates}, {Watson}, {Balmaceda}, {DeLuca}, \& {Martens}}]{Munoj2015}
{Mu{\~n}oz-Jaramillo}, A., {Senkpeil}, R.~R., {Windmueller}, J.~C., {et~al.} 2015, \bibinfo{title}{{Small-scale and Global Dynamos and the Area and Flux Distributions of Active Regions, Sunspot Groups, and Sunspots: A Multi-database Study},} \apj, 800, 48, \dodoi{10.1088/0004-637X/800/1/48}

\bibitem[{T. pandas~development team(2020)pandas~development team}]{pandas2020}
pandas~development team, T. 2020, \bibinfo{title}{pandas-dev/pandas: Pandas,}, latest Zenodo, \dodoi{10.5281/zenodo.3509134}

\bibitem[{A.~A. Pevtsov {et~al.}(2019)Pevtsov, Tlatova, Pevtsov, Heikkinen, Virtanen, Karachik, Bertello, Tlatov, Ulrich, \& Mursula}]{Pevtsov2019}
Pevtsov, A.~A., Tlatova, K.~A., Pevtsov, A.~A., {et~al.} 2019, \bibinfo{title}{Reconstructing Solar Magnetic Fields from Historical Observations: {{V}}. {{Sunspot}} Magnetic Field Measurements at {{Mount Wilson Observatory}},} Astronomy \& Astrophysics, 628, A103, \dodoi{10.1051/0004-6361/201834985}

\bibitem[{L. {Rightmire-Upton} {et~al.}(2012){Rightmire-Upton}, {Hathaway}, \& {Kosak}}]{RightmireUpton2012}
{Rightmire-Upton}, L., {Hathaway}, D.~H., \& {Kosak}, K. 2012, \bibinfo{title}{{Measurements of the Sun's High-latitude Meridional Circulation},} \apjl, 761, L14, \dodoi{10.1088/2041-8205/761/1/L14}

\bibitem[{C.~J. {Schrijver} {et~al.}(2002){Schrijver}, {De Rosa}, \& {Title}}]{Schrijver2002}
{Schrijver}, C.~J., {De Rosa}, M.~L., \& {Title}, A.~M. 2002, \bibinfo{title}{{What Is Missing from Our Understanding of Long-Term Solar and Heliospheric Activity?},} \apj, 577, 1006, \dodoi{10.1086/342247}

\bibitem[{C.~J. {Schrijver} \& A.~M. {Title}(2001){Schrijver} \& {Title}}]{Schrijver2001}
{Schrijver}, C.~J., \& {Title}, A.~M. 2001, \bibinfo{title}{{On the Formation of Polar Spots in Sun-like Stars},} \apj, 551, 1099, \dodoi{10.1086/320237}

\bibitem[{N.~R. {Sheeley}(1966){Sheeley}}]{Sheeley1966}
{Sheeley}, Jr., N.~R. 1966, \bibinfo{title}{{Measurements of Solar Magnetic Fields},} \apj, 144, 723, \dodoi{10.1086/148651}

\bibitem[{N.~R. {Sheeley}(1992){Sheeley}}]{Sheeley1992}
{Sheeley}, Jr., N.~R. 1992, in Astronomical Society of the Pacific Conference Series, Vol.~27, The Solar Cycle, ed. K.~L. {Harvey}, 1

\bibitem[{A. {Sreedevi} {et~al.}(2023){Sreedevi}, {Jha}, {Karak}, \& {Banerjee}}]{Sreedevi2023}
{Sreedevi}, A., {Jha}, B.~K., {Karak}, B.~B., \& {Banerjee}, D. 2023, \bibinfo{title}{{AutoTAB: Automatic Tracking Algorithm for Bipolar Magnetic Regions},} \apjs, 268, 58, \dodoi{10.3847/1538-4365/acec47}

\bibitem[{A. Sreedevi {et~al.}(2024)Sreedevi, Jha, Karak, \& Banerjee}]{sreedevi_analysis_2024}
Sreedevi, A., Jha, B.~K., Karak, B.~B., \& Banerjee, D. 2024, \bibinfo{title}{Analysis of {BMR} {Tilt} from {AutoTAB} {Catalog}: {Hinting} toward the {Thin} {Flux} {Tube} {Model}?} The Astrophysical Journal, 966, 112, \dodoi{10.3847/1538-4357/ad34b8}

\bibitem[{J.~O. {Stenflo} \& A.~G. {Kosovichev}(2012){Stenflo} \& {Kosovichev}}]{Stenflo2012a}
{Stenflo}, J.~O., \& {Kosovichev}, A.~G. 2012, \bibinfo{title}{{Bipolar Magnetic Regions on the Sun: Global Analysis of the SOHO/MDI Data Set},} \apj, 745, 129, \dodoi{10.1088/0004-637X/745/2/129}

\bibitem[{L. {Upton} \& D.~H. {Hathaway}(2014{\natexlab{a}}){Upton} \& {Hathaway}}]{Upton2014}
{Upton}, L., \& {Hathaway}, D.~H. 2014{\natexlab{a}}, \bibinfo{title}{{Predicting the Sun's Polar Magnetic Fields with a Surface Flux Transport Model},} \apj, 780, 5, \dodoi{10.1088/0004-637X/780/1/5}

\bibitem[{L. {Upton} \& D.~H. {Hathaway}(2014{\natexlab{b}}){Upton} \& {Hathaway}}]{Upton2014a}
{Upton}, L., \& {Hathaway}, D.~H. 2014{\natexlab{b}}, \bibinfo{title}{{Effects of Meridional Flow Variations on Solar Cycles 23 and 24},} \apj, 792, 142, \dodoi{10.1088/0004-637X/792/2/142}

\bibitem[{L. {Upton} \& B.~K. {Jha}(2025, In Prep.){Upton} \& {Jha}}]{Upton2025_are}
{Upton}, L., \& {Jha}, B.~K. 2025, In Prep., \bibinfo{title}{"Characterizing Solar Active Region Emergence over 100 Years of Sunspot Group Records",} In Preparation

\bibitem[{L.~A. Upton \& D.~H. Hathaway(2023)Upton \& Hathaway}]{Upton2023}
Upton, L.~A., \& Hathaway, D.~H. 2023, \bibinfo{title}{Solar Cycle Precursors and the Outlook for Cycle 25,} Journal of Geophysical Research: Space Physics, 128, e2023JA031681, \dodoi{https://doi.org/10.1029/2023JA031681}

\bibitem[{L.~A. {Upton} {et~al.}(2024){Upton}, {Ugarte-Urra}, {Warren}, \& {Hathaway}}]{Upton2024A}
{Upton}, L.~A., {Ugarte-Urra}, I., {Warren}, H.~P., \& {Hathaway}, D.~H. 2024, \bibinfo{title}{{The Advective Flux Transport Model: Improving the Far Side with Active Regions Observed by STEREO 304 {\r{A}}},} \apj, 968, 114, \dodoi{10.3847/1538-4357/ad40a5}

\bibitem[{I.~O.~I. Virtanen {et~al.}(2022)Virtanen, Pevtsov, Bertello, \& Mursula}]{Virtanen2022}
Virtanen, I. O.~I., Pevtsov, A.~A., Bertello, L., \& Mursula, K. 2022, \bibinfo{title}{Reconstructing Solar Magnetic Fields from Historical Observations: {{IX}}. {{The}} Photospheric Magnetic Field from 1915 to 1985,} Astronomy \& Astrophysics, 667, A168, \dodoi{10.1051/0004-6361/202244372}

\bibitem[{I.~O.~I. Virtanen {et~al.}(2019)Virtanen, Virtanen, Pevtsov, Bertello, Yeates, \& Mursula}]{Virtanen2019a}
Virtanen, I. O.~I., Virtanen, I.~I., Pevtsov, A.~A., {et~al.} 2019, \bibinfo{title}{Reconstructing Solar Magnetic Fields from Historical Observations: {{IV}}. {{Testing}} the Reconstruction Method,} Astronomy \& Astrophysics, 627, A11, \dodoi{10.1051/0004-6361/201935606}

\bibitem[{Y.~M. {Wang} \& J.~L. {Lean}(2021){Wang} \& {Lean}}]{Wang2021}
{Wang}, Y.~M., \& {Lean}, J.~L. 2021, \bibinfo{title}{{A New Reconstruction of the Sun's Magnetic Field and Total Irradiance since 1700},} \apj, 920, 100, \dodoi{10.3847/1538-4357/ac1740}

\bibitem[{Y.~M. {Wang} {et~al.}(2005){Wang}, {Lean}, \& {Sheeley}}]{Wang2005}
{Wang}, Y.~M., {Lean}, J.~L., \& {Sheeley}, Jr., N.~R. 2005, \bibinfo{title}{{Modeling the Sun's Magnetic Field and Irradiance since 1713},} \apj, 625, 522, \dodoi{10.1086/429689}

\bibitem[{Y.~M. {Wang} \& J. {Sheeley}(1991){Wang} \& {Sheeley}}]{Wang1991a}
{Wang}, Y.~M., \& {Sheeley}, N.~R., J. 1991, \bibinfo{title}{{Magnetic Flux Transport and the Sun's Dipole Moment: New Twists to the Babcock-Leighton Model},} \apj, 375, 761, \dodoi{10.1086/170240}

\bibitem[{Y.~M. {Wang} \& N.~R. {Sheeley}(1989){Wang} \& {Sheeley}}]{Wang1989}
{Wang}, Y.~M., \& {Sheeley}, Jr., N.~R. 1989, \bibinfo{title}{{Average Properties of Bipolar Magnetic Regions during Sunspot Cycle-21},} \solphys, 124, 81, \dodoi{10.1007/BF00146521}

\bibitem[{A.~R. {Yeates} {et~al.}(2025){Yeates}, {Bertello}, {Pevtsov}, \& {Pevtsov}}]{Yeates2025}
{Yeates}, A.~R., {Bertello}, L., {Pevtsov}, A.~A., \& {Pevtsov}, A.~A. 2025, \bibinfo{title}{{Latitude Quenching Nonlinearity in the Solar Dynamo},} \apj, 978, 147, \dodoi{10.3847/1538-4357/ad99d0}

\bibitem[{A.~R. {Yeates} {et~al.}(2023){Yeates}, {Cheung}, {Jiang}, {Petrovay}, \& {Wang}}]{Yeates2023}
{Yeates}, A.~R., {Cheung}, M. C.~M., {Jiang}, J., {Petrovay}, K., \& {Wang}, Y.-M. 2023, \bibinfo{title}{{Surface Flux Transport on the Sun},} \ssr, 219, 31, \dodoi{10.1007/s11214-023-00978-8}

\end{thebibliography}
\bibliographystyle{aasjournal}



\end{document}